\documentclass[table,xcdraw,smallextended]{svjour3}
\usepackage[utf8]{inputenc}
\makeatletter
\newsavebox\zb@x
\newcounter{z@@m}
\usepackage{calc}
\newdimen\B@r\newdimen\P@r
\newdimen\@zw\newdimen\@zh\newdimen\@zd

\newcommand{\zoombox}[2][0]{%
  \leavevmode%
  \sbox\zb@x{#2}%
  \setlength\B@r{1pt*\ratio{\wd\zb@x}{\ht\zb@x+\dp\zb@x}}%
  \setlength\P@r{1pt*\ratio{\paperwidth}{\paperheight}}%
  \ifdim\B@r>\P@r\relax%
    \setlength\@zw{\wd\zb@x}\setlength\@zh{\@zw*\ratio{\paperheight}{\paperwidth}}%
    \setlength\@zd{(\@zh-\ht\zb@x-\dp\zb@x)*\real{0.5}+\dp\zb@x}%
    \setlength\@zh{\@zh-\@zd}%
  \else%
    \setlength\@zh{\ht\zb@x+\dp\zb@x}%
    \setlength\@zw{\@zh*\ratio{\paperwidth}{\paperheight}}%
    \setlength\@zh{\ht\zb@x}\setlength\@zd{\dp\zb@x}%
  \fi%
  \makebox[0pt][l]{\makebox[\wd\zb@x][c]{\makebox[\@zw][l]{%
    \pdfdest name {zbfs\thez@@m} fitr
      width  \@zw\space
      height \@zh\space
      depth  \@zd\space
  }}}%
  \pdfdest name {zb\thez@@m} fitr
    width  \wd\zb@x\space
    height \ht\zb@x\space
    depth  \dp\zb@x\space
  \immediate\pdfannot 
    width  \wd\zb@x\space
    height \ht\zb@x\space
    depth  \dp\zb@x\space
  {%
    /Subtype/Link/H/N
    /Border [0 0 #1 [1 2]]
    /A <<
      /S/JavaScript
      /JS (
        if(typeof(zoomed)=='undefined'||!zoomed){
          var lastView=this.viewState;
          if(app.fs.isFullScreen) this.gotoNamedDest('zbfs\thez@@m');
          else this.gotoNamedDest('zb\thez@@m');
          zoomed=true;
        }else{
          this.viewState=lastView;
          zoomed=false;
        }
      )
    >>
  }%
  \usebox{\zb@x}%
  \stepcounter{z@@m}%
}   
\makeatother
\usepackage[hyphens]{url}
\usepackage{hyperref}

\usepackage[nounderscore]{syntax} 
\usepackage{color}
\usepackage{listings}
\usepackage{graphicx}

\usepackage[linesnumbered]{algorithm2e}
\usepackage[noend]{algpseudocode}      
\usepackage{paralist}
\usepackage{amsmath}
\usepackage{amssymb}
\usepackage{subfigure}
\usepackage{bchart}        
\usepackage{numprint}         
\usepackage{xspace}   
\usepackage{mdwlist}   
\usepackage{enumerate}  
\usepackage[]{tabularx}  
\usepackage[]{mdframed}
\usepackage{mathtools, cuted}

\usepackage{framed}
\usepackage{newfloat} 

\usepackage{xfrac}

\usepackage{colortbl}

\usepackage{tikz}
\usepackage[ligature, inference]{semantic}

\usepackage{paralist}

\usepackage{wrapfig}

\definecolor{pblue}{rgb}{0.13,0.13,1}
\definecolor{pgreen}{rgb}{0,0.5,0}
\definecolor{pred}{rgb}{0.9,0,0}
\definecolor{pgrey}{rgb}{0.46,0.45,0.48}

\definecolor{darkgreen}{rgb}{0,0.45,0}

\usepackage{xspace}

\newcommand{\modified}[1]{{\color{black}#1}} 
\newcommand{\nmodified}[1]{{\color{darkgreen}#1}} 

\newcommand{\add}{{\sc add}\xspace}
\newcommand{\delete}{{\sc delete}\xspace}
\newcommand{\replace}{{\sc replace}\xspace}
\newcommand{\addmi}{{\sc add method invocation}\xspace}
\newcommand{\swap}{{\sc swap subtype}\xspace}
\newcommand{\loopflip}{{\sc loop flip}\xspace}

\newcommand{\transf}{program transformation\xspace}
\newcommand{\transfs}{program transformations\xspace}

\newcommand{\ctransfs}{Program transformations\xspace}

\mathlig{->}{\rightarrow}

\title{
A Journey Among Java Neutral Program Variants
 \thanks{We would like to thank Amine Benelallam  and Paul Bettega for their feedback on this work.
This work has been partially supported by the EU Project STAMP ICT-16-10 No.731529 under the H2020 framework program, by the Wallenberg Autonomous Systems and Software Program, by the TrustFull project financed by the Swedish Foundation for Strategic Research and by the OSS-Orange-Inria project.}
}

\author{Nicolas Harrand         \and
        Simon Allier \and
        Marcelino Rodriguez-Cancio \and
        Martin Monperrus \and
        Benoit Baudry
}

\institute{
Nicolas Harrand \at KTH, Sweden \\
\email{harrand@kth.se}          
\and
Simon Allier \at DGA, France \\
\email{simon.allier@intradef.gouv.fr}                      
\and
Marcelino Rodiguez-Cancio \at Vanderbildt University, USA \\
\email{marcelino.rguez.cancio@gmail.com}  
\and
Martin Monperrus \at KTH, Sweden \\
\email{martin.monperrus@csc.kth.se}           
\and
Benoit Baudry \at KTH, Sweden \\
\email{baudry@kth.se}                 
}

\date{Received: date / Accepted: date}

\begin{document}

\maketitle

\begin{abstract}
\modified{\emph{Neutral program variants} are alternative implementations of a program, yet equivalent  with respect to the test suite.}
Techniques such as approximate computing or genetic improvement share the intuition that potential for enhancements lies in these acceptable behavioral differences (e.g., enhanced performance or  reliability). Yet, the automatic synthesis of neutral program variants, through \emph{\transfs}  remains a key challenge.

This work aims at characterizing \emph{plastic code regions} in Java programs, i.e., the \modified{code regions that are modifiable while maintaining functional correctness, according to a test suite}.
Our empirical study relies on automatic variations of  6 real-world Java programs. First, we transform these programs with three state-of-the-art \transfs: add, replace and delete statements. We get a pool of 23445 neutral variants, from which we gather the following novel insights: developers naturally write code that supports fine-grain behavioral changes; statement deletion is a surprisingly effective \transf; high-level design decisions, such as the choice of a data structure, are natural points that can evolve while keeping functionality. 

Second, we design 3 novel \transfs, targeted at specific plastic regions. New experiments reveal that respectively 60\%, 58\% and 73\% of the synthesized variants (175688 in total) are neutral and exhibit execution traces that are different from the original.
\end{abstract} 

\section{Introduction}

Neutral program variants are at the core of automatic software enhancement. 
The intuition is that these variants that are  different from the original, yet are similar have the potential for enhanced performance, security or resilience.
Approximate computing explores how program variants can provide different trade-offs between accuracy and resource consumption \cite{mittal2016survey}. 
Software diversity aims at using these variants to reduce the knowledge that an attacker can take for granted when designing exploits \cite{baudry2015}. 
Genetic improvement \cite{petke2017genetic} automatically searches the space of program variants for improved performance. 

Despite their key role, the automatic synthesis of neutral program variants, is still a major challenge \modified{because of the size of the search space. Starting from one initial program that one aim to improve, there exists a vast amount of possible variants that can be synthesized through small code transformations, most of which do not compile or do not pass the test suite (i.e., ill-formed variants). Exploring this search space randomly can produce a large number of ill-formed variants that are useless for automatic improvement, but still require resources to synthesize and try to compile and test. Our work aims at reducing the number of ill-formed variants that are generated while exploring the space of program variants for automatic improvement tasks.} We focus on two specific challenges: understanding \emph{how} and \emph{where} to transform a program to synthesize a neutral variant. The \emph{how} part refers to the design of \emph{\transfs} that introduce some behavioral variations.  The \emph{where} refers to  the parts of a program that can stand behavioral variations, while maintaining the overall functionality similar to the original program. 
We call these parts of programs the \emph{plastic code regions}. \modified{With the term ``plastic'' we want to capture a specific characteristic of certain code regions: their ``malleability'', or they intrinsic capability at being changed to another code while keeping functional correctness, with respect to a given test suite. If we can identify such code regions, they become natural candidates for transformations that aim to synthesize neutral variants. This concept of plastic region is close to the concept of forgiving code regions explored by Rinard \cite{rinard2012} or of mutational robustness explored by Schulte \cite{Schulte13}. The conceptual difference is that Rinard and Schulte reason about the ability to tolerate perturbations, while, with the term ``plastic'', we aim at characterizing the ability of the code to exist in multiple forms.}

Our work aims to characterize these plastic code regions. 
This journey focuses on  Java programs  and the in-depth analysis of various \transfs on  6 large, mature, open source Java projects. We articulate our journey around three main parts. 
First, we run state of the art \transfs \cite{Baudry14,Schulte13} that add, delete or replace an AST node. 
We consider that a transformation synthesizes a \emph{neutral variant} if the variant compiles and successfully passes the test suite of the original program. 
This first contribution is a conceptual replication \cite{shull_role_2008} of the work by Schulte and colleagues \cite{Schulte13}. 
This replication addresses two threats to the validity of Schulte's results: our methodology mitigates internal threats, by using another tool to detect neutral variants, and our experiment mitigates external threats by experimenting with a new set of programs, in a different programming language.

Second, we analyze a set of 23445 neutral variants.  We provide a quantitative analysis of the types of AST nodes and the types of transformations that more likely yield neutral variants. We analyze the interplay between the synthesis of neutral program variants and the specification of the original program provided as a set of test cases. Also, we manually analyze dozens of neutral variants to provide a qualitative analysis of plastic code regions and the role they play in Java programs.

In the third part of our investigation, we design and experiment with three novel, targeted \transfs: \addmi, \swap and \loopflip. Our experiments with our 6 Java projects demonstrate a significant increase in the rate of neutral variants among the program variants (respectively 60\%, 58\% and 73\%). We consolidate these results by assessing that the neutral variants indeed implement behavior differences: we trace the execution of these variants, and observe that all neutral variants actually exhibit behavior diversity.

In summary, this work contributes novel insights about neutral program variants, as follows:
\begin{itemize}
    \item A conceptual replication of the work by Schulte and colleagues \cite{Schulte13} about the existence of neutral variants, with a new tool, new study subjects and a different programming language
	\item A large scale quantitative analysis of the types of Java language constructs that are prone to neutral variants synthesis with the state of the art \transfs: add, delete and replace AST nodes
	\item  A deep, qualitative analysis of plastic code regions that can be exploited to design efficient \transfs 
	\item Three targeted \transfs that significantly increase the ratio of neutral variants, compared to the state of the art
    \item Open tools and datasets to support the reproduction of the experiments, available at: \url{https://github.com/castor-software/journey-paper-replication}
\end{itemize}

The rest of this paper is organized as follows. In \autoref{sec:background}, we define the terminology for this work. In \autoref{sec:exp-protocol}, we introduce  the experimental protocol that we follow in order to investigate the synthesis of neutral variants. 
In \autoref{sec:RQ1}, we analyze the types of \transfs and AST nodes that more likely yield neutral variants. In \autoref{sec:taxonomy}, we manually explore and categorize neutral variants according to the role of the code region that has been transformed. In \autoref{sec:targeted}, we leverage the analysis of previous sections to design novel \transfs targeted at specific code regions. 
In \autoref{sec:discussion}, we discuss some key findings of this study. \autoref{sec:threats} elaborates on the threats to the validity of this work, \autoref{sec:related} discusses related work and we conclude in \autoref{sec:conclusion}.

\section{Background and Definitions}
\label{sec:background}

Here we define the key concepts that we leverage to explore the different regions of Java programs that are prone to the synthesis of neutral program variants.

\subsection{Generic \transfs}

\begin{figure}[ht]
 \subfigure[Original]{
			\includegraphics[width=0.22\columnwidth]{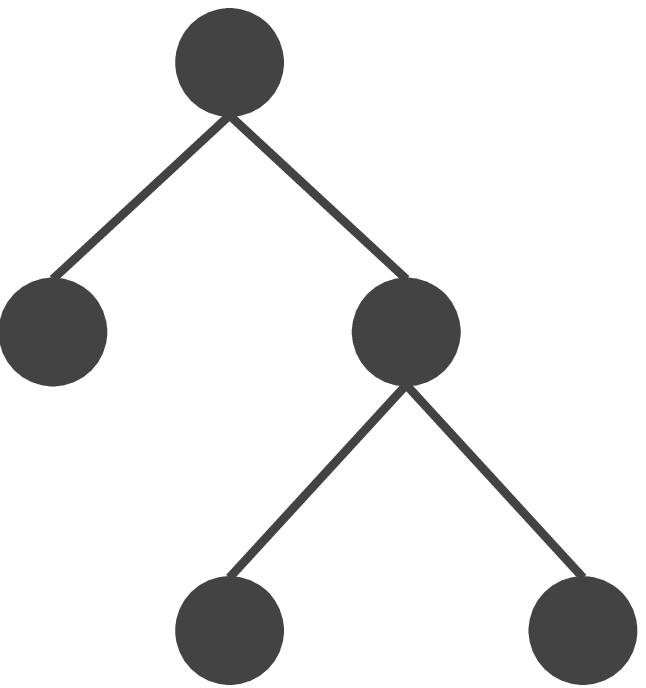}
			\label{fig:original}
 }
 \subfigure[\add]{
			\includegraphics[width=0.22\columnwidth]{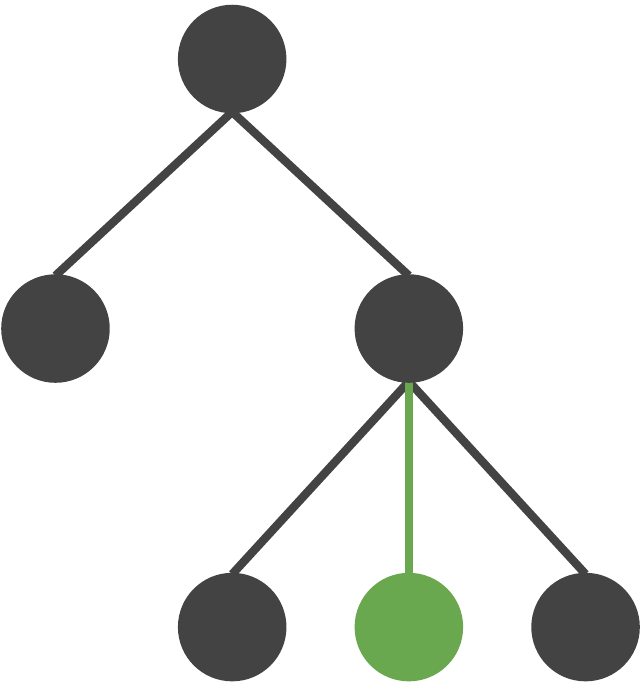}
			\label{fig:addop}
 }
 \subfigure[\delete]{
			\includegraphics[width=0.22\columnwidth]{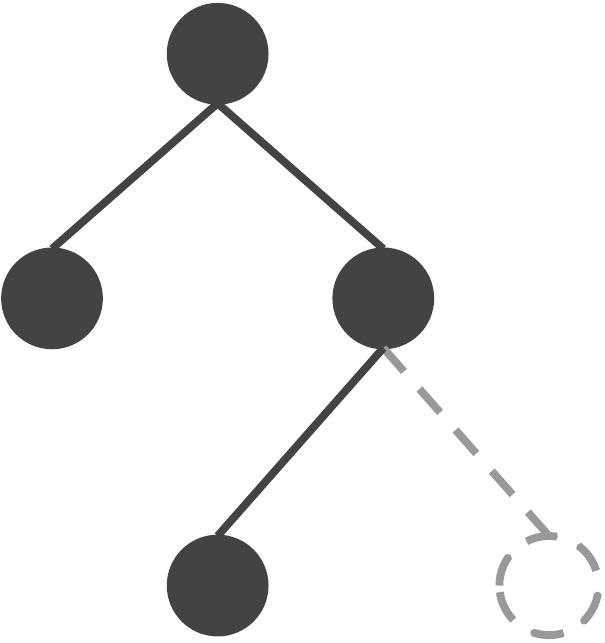}
			\label{fig:deleteop}
 }
 \subfigure[\replace]{
			\includegraphics[width=0.22\columnwidth]{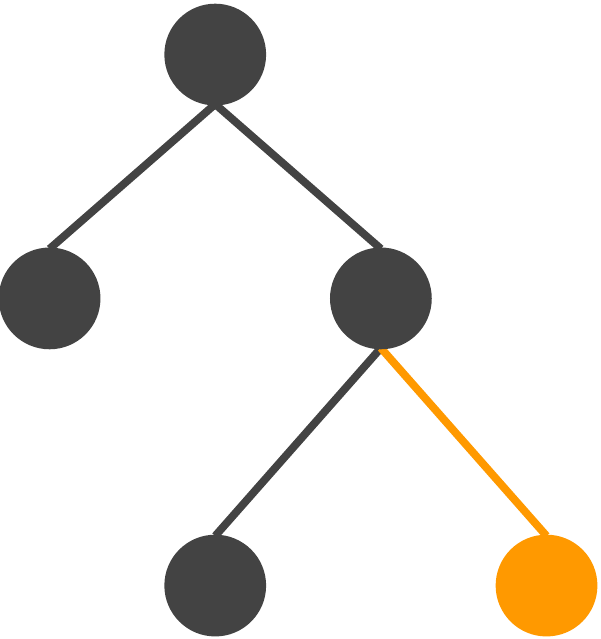}
			\label{fig:replaceop}
 }
 \caption{Generic \transfs}
 \label{fig:generics}
 \end{figure}

Given an initial program, which comes along with a test suite, we consider three generic \transfs on source code that have been defined in previous work \cite{Baudry14,petke2017genetic,Schulte13}.
These transformations operate on the abstract syntax tree (AST).  \nmodified{In this context we call \textit{code region} a sub tree present in a program AST.} 
First, we randomly select a statement node in the AST, we check if it is covered by one test case at least (to prevent transforming dead code), then, we consider three types of transformations (cf. \autoref{fig:generics}).
\begin{definition}\label{def:transfo} \textbf{\transfs}.
We consider the following three transformations on AST nodes
	\begin{itemize}
		\item delete the node and the subsequent subtree (\delete, Figure \ref{fig:deleteop});
		\item add a node just before the selected one (\add, Figure \ref{fig:addop});
		\item replace the node and the  subtree by another one (\replace, Figure \ref{fig:replaceop}). 	 
	\end{itemize}
\end{definition}

\modified{
\begin{definition}\label{def:location} \textbf{Location}. The statement at which we perform a \transf is called the location.
\end{definition}
}
\begin{definition}\label{def:transplant} \textbf{Transplant}. For \add and \replace, the statement that is copied and inserted is called  the transplant statement. 
\end{definition}

\modified{This terminology (\autoref{def:location} and \autoref{def:transplant}) follow a convention established by Barr and colleagues\cite{barr2015automated}.}

We add further constraints to the generic \transfs in order to increase the chance of synthesizing neutral variants. For \add and \replace, we consider transplant statements from the same program as the location (we do not synthesize new code, nor take code from other programs). 
We also consider the following two additional steps :
\begin{itemize}
	\item We build the type signature of the location: the list of all variable types that are used in the location and the return type of the statement.  The transplant shall be randomly selected only among statements that have a compatible signature.
	\item When injecting the transplant (as a replacement or an addition to the transplant), the variables of the statement are renamed with names of variables of the same types that are in the scope of the location. \modified{Similar restrictions are common in the GI literature, for example Yuan and colleagues use a type matching based approach\cite{yuan2018arja}}.
\end{itemize}

\autoref{fig:tp} shows an excerpt of program, in which we have selected one location. \autoref{fig:transplant} is a transplant example, i.e., an existing statement extracted from the same program. In order to insert the transplant at the location, we need to rename the variables with names that fit the namespace. The expression \texttt{inAvail < max} can be rewritten in 4 different ways: each integer variable can be replaced by one of the two integer variable identifiers (\texttt{a} or \texttt{i}). The statement \texttt{context.eof = true;} can be rewritten in one single way, rewriting \texttt{context.eof} into \texttt{b}.

 \begin{figure}[ht]
 \begin{minipage}[b]{0.75\linewidth}
 \centering
        \scriptsize
      \begin{lstlisting}[language=java]
if (inAvail < max) {
  context.eof = true;
} 
\end{lstlisting}
    \caption{\label{fig:transplant}Transplant}
 \end{minipage}
 \hspace{0.5cm}
 \begin{minipage}[b]{0.5\linewidth}
 \centering
      \scriptsize
      \begin{lstlisting}[language=java]
class A {
  int i = 0;
  void m(int a) {
    boolean b = false;
    [...]
    //Location
    [...]	
  }
}     \end{lstlisting}
 \caption{\label{fig:tp}Location}
 \end{minipage}
 \begin{minipage}[b]{0.5\linewidth}
 \centering
      \scriptsize
      \begin{lstlisting}[language=java]
class A {
  int i = 0;
  void m(int a) {
    boolean b = false;
    [...]
    if (this.i < a) {
      b = true;
    } 
    [...]	
  }
}
     \end{lstlisting}
 \caption{\label{fig:transformed}Transformed code}
 \end{minipage}
 \end{figure}

There are different reasons for which a random add or replace fails at producing a compilable variant. 
Hence we introduce different preconditions to limit the number of meaningless variants.

For \replace, we enforce that: a statement cannot be replaced by itself; for both \add and \replace, statements of type \emph{case}, AST nodes of type \emph{variable instantiation}, \emph{return}, \emph{throw} are only replaced by statements of the same type; the type of returned value in a \emph{return} statement must be the same for the original and for its replacement.

\subsection{Neutral variant}
\label{def:sosiefication}

Given a program $P$ and a test suite $TS$ for $P$, a \transf can  synthesize a variant program $\tau(P)$, which  falls into one of the following categories: (i) $\tau(P)$ does not compile; (ii) the variant compiles but does not pass all the tests in $TS$: $\exists t \in TS | fail(t,\tau(P))$; (iii) the variant compiles and passes the same test suite as the original program: $\forall t \in TS | pass(t,\tau(P))$. This work focuses on the latter category, i.e., all variants that are equivalent to the original modulo the test suite. 
We call such variants \emph{neutral variants}.

\begin{definition}\label{def:sosie} \textbf{Neutral variant}.
	Given a program $P$, a test suite $TS$ for $P$ and a program transformation $\tau$, a variant $\tau(P)$ is a neutral variant of $P$ if the two following conditions hold
	1) $\tau(P)$ results from a \transf on a region of $P$ that is covered by at least one test case of $TS$;
	2) $\forall t \in TS | pass(t,\tau(P))$
\end{definition}

This work aims at characterizing the code regions of Java programs where \transfs  are the most likely to synthesize neutral variants.

\section{Experimental protocol}
\label{sec:exp-protocol}

\ctransfs are instrumental for automatic software improvement, and code plasticity is the property of software that supports these transformations. In what follows, we design a protocol to analyze the interplay between transformations, the programming language and code plasticity.

\subsection{Protocol}
\label{sec:protocol}

In this paper, we perform the following experiment.

The experiment is budget-based: we try neither to exhaustively visit the search space nor to have a fixed-size sample. 
Since the investigation of neutral variants is an  expensive process, our computation platform is Grid5000, a scientific platform for parallel, large-scale computation \cite{bolze2006grid}.
\nmodified{We submit one batch of single \transfs for each program that is run as long as resources (CPU and memory) are available on the grid. Both locations and transplant are selected randomly within the rules detailed in \autoref{sec:background}.}
\modified{Then, for each variant that compiles, we extract or compute the metrics described in \autoref{sec:metrics}.}
We  also manually analyze dozens of neutral variants in order to build a taxonomy of plastic code regions.

In the second part of our study, we refine the \transfs defined above, in order to target specific code regions. We run another round of experiments to determine the impact of targeted transformations on the neutral variant rate. 

\subsection{Dataset}

\begin{table}[ht]
	\centering
	\caption{Descriptive statistics about our subject programs}
	\begin{tabularx}{\columnwidth}{lXXXX}
		\hline
		& \#classes & \#stmt & \#TC & cov.   \\
		\hline
		commons-lang 3.3.2 & 132 & 8442 & 2514 & 94\% \\
		commons-collections 4.0  & 286 & 6780 & 13677 & 84\% \\
		commons-codec 1.10  & 60 & 2695 & 662 & 96\% \\
		commons-io 2.4 & 103 & 2573 & 966 &  87\% \\
		Gson 2.4 & 66 & 2377 & 966 &  79\% \\
		jgit 3.7.0 & 666 & 22333 & 3341 & 70\%\\
		\hline
	\end{tabularx}
	\label{tab:subjects}
\end{table}

We consider the 6 programs presented in \autoref{tab:subjects}. \modified{They were manually selected among popular java programs (cf \footnote{\url{mvnrepository.com}}) with a strong test suite.}
All  programs are popular Java libraries developed by either the Apache foundation, Google or Eclipse.\footnote{\scriptsize The exact versions of the library and the whole dataset is available here: \url{https://github.com/castor-software/journey-paper-replication/tree/master/projects}}
The  second column gives the number of classes,
the third column the number of statements. This latter number approximates the size of the search space for our \transfs.
Column 4 provides the number of test case executions when running the test suite and column 5 gives the statement coverage rate. (This number of test case execution corresponds to the number of Junit test methods as reported by maven.)

The programs range between 60 and 666 classes. 
All of them are tested with very large test suites that include hundreds of test cases executing the program in many different situations. 
One can notice the extremely high number of test cases executed on commons-collection. 
This results from an extensive usage of inheritance in the test suite, hence many test cases are executed multiple times (e.g., test cases that test methods declared in abstract classes). 
The test suites cover most of the program (up to 96\% statement coverage for commons-codec). Jgit is the exception (only 70\% coverage): it includes many classes meant to connect to different remote git servers, which are not covered by the unit test cases (due to the difficulty of stubbing these servers).
This dataset provides a solid basis to investigate the role plastic code regions play to produce modulo-test equivalent program variants.

\subsection{Metrics}
\label{sec:metrics}

\modified{
\begin{definition}\label{def:SR}  \textbf{Neutral Variant Rate (NVR)}
	is the ratio between the number of neutral variants and the number of transformations that produce a variant that compiles: $\# NeutralVariants/\# Compile$.
\end{definition}
}

The neutral variant rate is a key metric to capture the plasticity of a code region: the higher it is for a certain region, the more this region can be used by \transfs to synthesize valid variants. 
\modified{It is designed to consider only variants that compile, because (i) our goal is to study what characteristics impact a program tolerance for alternative implementation (ii) we compare it for transformations with widely different $\#CompileVariants/\# Transformation$ ratios. This ratio is more linked to the transformation implementation than to whatever characteristic of the targeted program region. It is noteworthy that running tests is the actual costly part of the search for neutral variants. Non-compilable variants fail fast and therefore, do not cost much search time.}

We collect the following metrics to characterize the regions where we perform \transfs.

\modified{\begin{definition}\label{def:features}   
	\textbf{Location features}:    
	Let us call $loc$ the location yielding the neutral variant. We focus on the following features:
	\begin{inparaenum}[\itshape 1\upshape)]
		\item$TC_{loc}$ is the number of test cases that execute $loc$.
		\item$Transfo_{loc}$ is a categorical feature that characterizes the type of transformation that we performed on $loc$: \add,  \delete or  \replace. This can be further refined by considering the type of AST node where the transformation occurs.
	\end{inparaenum}
\end{definition}}

\subsection{Research Questions}

Our journey among neutral variants is organized around the following research questions:

\textbf{RQ1. To what extent can we generate neutral variants through random \transfs?}

This first question can be seen as a conceptual replication of Schulte and colleagues' \cite{Schulte13}'s experiment demonstrating software mutational robustness. Here, we analyze the same phenomenon with a new transformation tool, new study subjects and in a different programming language. 

\textbf{RQ2. To what extent does the number of test cases covering a certain region impact its ability to support \transfs?}

This question addresses the interplay between the synthesis of neutral variants and the specification for specific code regions. Since our notion of neutral variant is modulo-test, we  check if the number of test cases that cover the location influences the ability to synthesize a neutral variant.

\textbf{RQ3. Are all program regions equally prone to produce neutral variants under \transfs?}

In this question, we are interested in analyzing whether the type of AST node or the type of transformation has an impact on the neutral variant rate. 
For instance, it may happen that loops are more plastic than assignments.
We study three dimensions in the qualification of transformations: 
1) how they are applied (addition of new code versus deletion of existing code);
2) where they are applied, i.e. the type of the locations (e.g. conditions versus method invocations);
and 
3) for \add and \replace, the type of the transplant.

\textbf{RQ4. What roles the code regions prone to neutral variant synthesis play in the program?} 

This question relies on a manual inquiry of dozens of neutral variants from all programs of our dataset, to build a taxonomy of program neutral variants. 
Here, we categorize different roles that certain code regions can play (e.g., optimization or data checking code) and relate this role to the plasticity of the region. 

\textbf{RQ5. Can \transfs target specific plastic code regions in order to increase their capacity at synthesizing neutral variants that exhibit behavioral variations?}

We exploit the insights gained in RQ3 and RQ4 to define novel types of \transfs, which refine the \add and  \replace generic transformations: \addmi, \swap, \loopflip. 
These transformations perform additional code analysis to select the location.  
\modified{This question investigates whether this refinement helps to reduce the number of variants that are not neutral program variants hence cannot be used as candidates for modulo test equivalent improvement.}

\subsection{Tools}
\label{sec:tools}

To conduct the  experiments described in this paper, we have implemented a tool that runs \transfs on Java programs and automatically runs a test suits on the variant, in order to select neutral variants. This tool, \textbf{Sosiefier} is open source and available online. \footnote{\url{https://github.com/DIVERSIFY-project/sosiefier}} The analysis and transformation of the JAVA AST mostly relies on another open source library called Spoon\cite{spoon}. 

To capture, align and compare execution traces described in \autoref{sec:targeted}, we have  implemented  \textbf{yajta}\footnote{\url{https://github.com/castor-software/yajta}}, a library to tailor runtime probes and trace representations. It uses a Java agent, which  instruments Java bytecode with Javassist \cite{chiba2000load}, to collect log information about the execution. 
Scalability is a key challenge here, since the insertion of probes on every branch of every method represents a considerable overhead both in terms of execution time, and heap size. For example,  a single test run can generate a trace up to GBs of data, which turns into a performance bottleneck when comparing the traces from hundreds of variants. This is especially true for performance test cases such as \texttt{PhoneticEnginePerformanceTest} (335 500 702 method calls and 990 617 578 branches executed) in \texttt{commons-codec}. These issues are well described in the work of Kim \textit{et al}.\cite{Kim2015}.

Consequently, we optimized the tracing process as follows:
\begin{inparaenum}[i)]
\item execute and compare only the test cases that actually cover the location in the original program ;
\item add transformation-specific knowledge to target the logs (e.g. the addition of a method invocation only requires to trace method call) ; and
\item collect and store complete traces  only for the original program, and compare this trace with the variant behavior  \emph{on-the-fly}. This way, we determine, at runtime, if a divergence occurs and we do not need to store the execution trace of the variant.
\end{inparaenum}

\section{Results}	

\subsection{Neutral variant rate of random transformations}
\label{sec:RQ1}This section focuses on RQ1.

\begin{framed}
	RQ1. To what extent can we generate neutral variants through random transformations?
\end{framed}

\begin{table}[ht]
	\centering
	\caption{Neutral variant rate for the synthesis of neutral program variants with the generic, random \transfs}
	\small{
		\begin{tabularx}{\columnwidth}{lXXXXX}
			\hline
			& add & del & rep & NVR & exploration  \\
			\hline

            commons-codec & 289 & 146 & 266 & 18.0\% & 91.9\% \\ 
            commons-collections & 3912 & 754 & 3960 & 21.8\% & 83.3\% \\
            commons-io & 1754 & 319 & 1472 & 21.1\% & 92\% \\
            commons-lang & 419 & 190 & 537 & 15.7\% & 78\% \\
            gson & 2199 & 215 & 1897 & 25.3\% & 80.3\% \\
            jgit & 1924 & 1375 & 2963 & 30.0\% & 57\% \\
            \hline
            total & 10078 & 2809 & 10558 & 23.9\% & - \\
            \hline
		\end{tabularx}
	}

	\label{tab:overall-results}
\end{table}

We run \transfs on our six case studies (cf. \autoref{tab:subjects}).
\autoref{tab:overall-results} gives the key data about the neutral variants computed with the budget-based approach described in \autoref{sec:protocol}.
It sums up the results of the 180 207 variants generated, from which 98225 compile and 23445 are neutral variants.
The second, third and fourth columns indicate the number of neutral variants synthesized by \add, \delete or \replace. 
The fifth column indicates the global neutral variant rate (NVR) as defined in \autoref{def:SR}, i.e., the rate of  neutral variants among  all variants that we generated and that compile.
The last column  (exploration) indicates  the rate of program statements  on which we ran a transformation, i.e., the extent to which we explored the space of locations. The low exploration  rate for \texttt{jgit} is related to the large size of the project: since our exploration of \transfs has a bounded resource budget, we could not cover a large program as much as a small one.

The data in \autoref{tab:overall-results} provides clear evidence that it is possible to synthesize neutral variants with \transfs.  In other words, it is possible to transform statements of programs and obtain programs that compile and are equivalent to the original, modulo the test suite. 

The program variants that compile are neutral variants in up to 30\% of the cases (for \texttt{jgit}).

This first research question is a conceptual replication of the study of Schulte and colleagues \cite{Schulte13}. Their \transfs are the same as ours. Yet, they ran experiments on a very different set of study subjects: 
 22 programs written in C, of size ranging from 34 to 59K lines of code and with test suites of various coverage ratios (from 100\% to coverage below 1\%). They also run experiments on the assembly counterpart of these programs.  
Their results show that 33.9\% of the variants of on C code are neutral, with a standard deviation of 10. They also obtain 39.6\% of neutral variants at the assembly level, with a standard deviation of 22 on assembly variants.

Our results confirm the main observation Schulte and colleagues: running \add, \delete and \replace randomly can synthesize a significant ratio of neutral program variants. 
The neutral variant rate between both our and Schulte's experiments are of the same order of magnitude. Their experiments generate slightly more neutral variants, which could indicate that different programming languages allow various degrees of plasticity. In particular, a stronger type system can limit code plasticity. Yet, the in-depth analysis of differences between languages is outside the scope of this paper.

\begin{framed}

Answer to RQ1: \ctransfs, applied in random code regions, can synthesize neutral program variants on Java source code. The ratio of neutral variants varies between 15.7\% and 30.0\%, out of thousands of variants, for our dataset. These new results confirm the main observations of Schulte and colleagues.
\end{framed}

\newpage
\subsection{Sensitivity to the test suite}
\label{sec:test}

\begin{framed}
RQ2. To what extent does the number of test cases covering a certain region impact its ability to support \transfs? 
\end{framed}

Here, we check if the number of test cases that cover a statement affects the plasticity that we observe. In other words, we evaluate the importance of the number of test cases that cover a location with respect to the probability of synthesizing a neutral variant when we transform that point with one of our \transfs.

\begin{figure}[ht]
	\centering
	\includegraphics[width=\columnwidth]{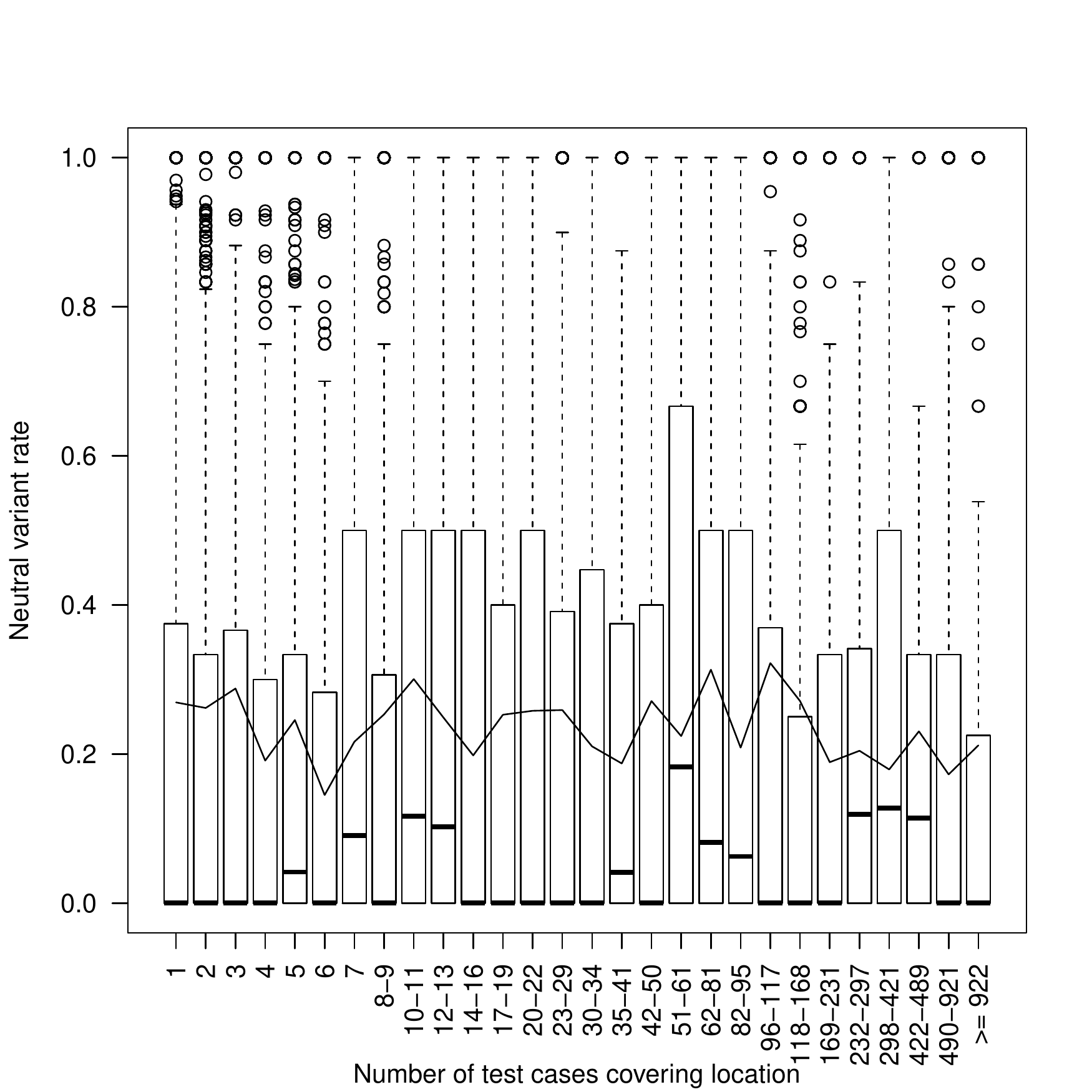}
	\caption{Neutral variant rate with respect to the number of test cases covering a location}
	\label{fig:sr-log-tc-all}
\end{figure}

In order to analyze this impact, we look at the distribution of neutral variant rate for all trials made on statements covered by a given number of test cases. 
Yet, in all projects, the distribution of statements according to the number of test cases that cover it is extremely skewed: more than half of the statements are covered by only one test case and then there is a long tail of few statements that are covered by tens and even hundreds of test cases.

\autoref{fig:sr-log-tc-all} represents the following information, given any location at which we synthesized one or multiple variants that compile, what is the probability that we succeed in getting a neutral variant, given the number of test cases that cover the location?
Because of the skewed distribution of statements with respect to the number of covering test cases, we group data in bins of locations that represent at least 4000 transformations.
Bins for low numbers of test cases cover a narrower range of values because statements covered by few tests are more common than statements covered by a large amounts of tests.

The broken line represents the average neutral variant rate per bin of locations. Boxes represent the first and last quartile and the median for the distribution of neutral variant rate for statements covered by $n$ test cases. Circles represents outliers (outside of a 95\% confidence interval) statement for each classes.
For example, for the 5943 locations covered by 1 test case, the weighted average neutral variant rate is 26.9\% and 25\% of these points support the synthesis of neutral variants in more than 37.5\% of the trials. Outliers are locations for which neutral variant rate is above 93.8\%.

For 17 out of 28 bins, the median neutral variant rate is 0\%, meaning that, for at least half of the locations, none of the variants tried are actually neutral.
Meanwhile, the first quartile is above 0\% for all bins. This means that we successfully synthesized neutral variants for at least 25\% of statements covered, independently of the amount of test cases (for 11 bins it is actually more than 50\% of statements).
The average neutral variant rate is close to the overall neutral variant rate of 23.9\%, whatever the number of test cases covering the location.

Under the assumption of a linear model, the part of the neutral variant rate explained by the number of test cases is negligible (Adjusted R-squared:  0.002036).
This implies either that the ability to synthesize a neutral variant on a given statement is not significantly influenced by the number of test cases that cover it with a linear model.

\begin{framed}
Answer to RQ2: the number of test cases that cover a location is independent from the ability to synthesize a neutral variant at this point. We believe that this indicates the presence of inherent \emph{code plasticity}, a concept  for which we propose a first characterization in the RQ5. 
\nmodified{To some extent, the neutral variant rate on locations that are covered by large numbers of test cases reflects this amount of software plasticity.} 
\end{framed}

\subsection{Language level plasticity}
\label{sec:nodes}

\begin{framed}
RQ3 Are all program regions equally prone to produce neutral variants under \transfs?
\end{framed}

\begin{table}[ht]
	\centering
	\caption{Distribution of Statement type across projects}
	\small{
		\begin{tabularx}{\columnwidth}{lXXX}
			\hline
			Node Type & Min & Med & Max\\
			\hline
			Invocation      & 34\%    & 37\%     & 39\%\\
			Assignment      & 17\%    & 19\%     & 22\%\\
			Return          & 10\%    & 13\%     & 19\%\\
			If              & 9.4\%   & 10\%     & 14\%\\
			ConstructorCall & 4.2\%   & 6.6\%    & 8.8\%\\
			UnaryOperator   & 3.1\%   & 3.8\%    & 8.6\%\\
			Throw           & 1.7\%   & 2.9\%    & 4.4\%\\
			Case            & 0.13\%  & 1.2\%    & 2.6\%\\
			For             & 0.55\%  & 0.76\%   & 1.5\%\\
			ForEach         & 0.37\%  & 0.72\%   & 0.87\%\\
			Try             & 0.17\%  & 0.65\%   & 1.4\%\\
			While           & 0.40\%  & 0.62\%   & 0.85\%\\
			Break           & 0.18\%  & 0.54\%   & 1.6\%\\
			Continue        & 0.018\% & 0.21\%   & 0.65\%\\
			Switch          & 0.033\% & 0.17\%   & 0.32\%\\
			Synchronized    & 0       & 0.048\%  & 0.21\%\\
			Enum            & 0       & 0.042\%  & 0.094\%\\
			Do              & 0       & 0.032\%  & 0.091\%\\
			Assert          & 0       & 2.68e-03\%  & 0.0014\%\\
			\hline
		\end{tabularx}
	}
	\label{tab:distrib-stmt}
\end{table}

As a preliminary step for our analysis of the plasticity of language structures, we analyze the usage frequency of each construct. 
\autoref{tab:distrib-stmt}, summarizes the usage distribution of each construct listed by decreasing median frequency. 
It appears that 6 constructs are frequently used, in approximately the same proportion in all projects (the top 6 lines of the table). There is no surprise here: these constructs correspond to the fundamental statements of any object-oriented program (\texttt{assignment}, \texttt{if}, \texttt{invocation}, \texttt{return}, \texttt{constructor call} and \texttt{unaryOperator}).

The 13 other constructs present in the table are an order of magnitude less frequent than the top constructs. They are also used in more various ways across programs. For instance, commons-collections favors \texttt{for-each} and \texttt{while} loops, while commons-codec uses \texttt{for} loops. This can be explained by the different types of structure that these projects use: collections vs arrays.

The use of \texttt{switch} and its child nodes (\texttt{break}, \texttt{case}, and \texttt{continue}) as well as \texttt{try} are also unequally distributed across projects.
This disparity partly explains the variation in the observations presented in the following section: uncommon constructs lead to more variations.

\subsubsection{\add}

\begin{figure}[ht]
	\centering
	\includegraphics[width=\columnwidth]{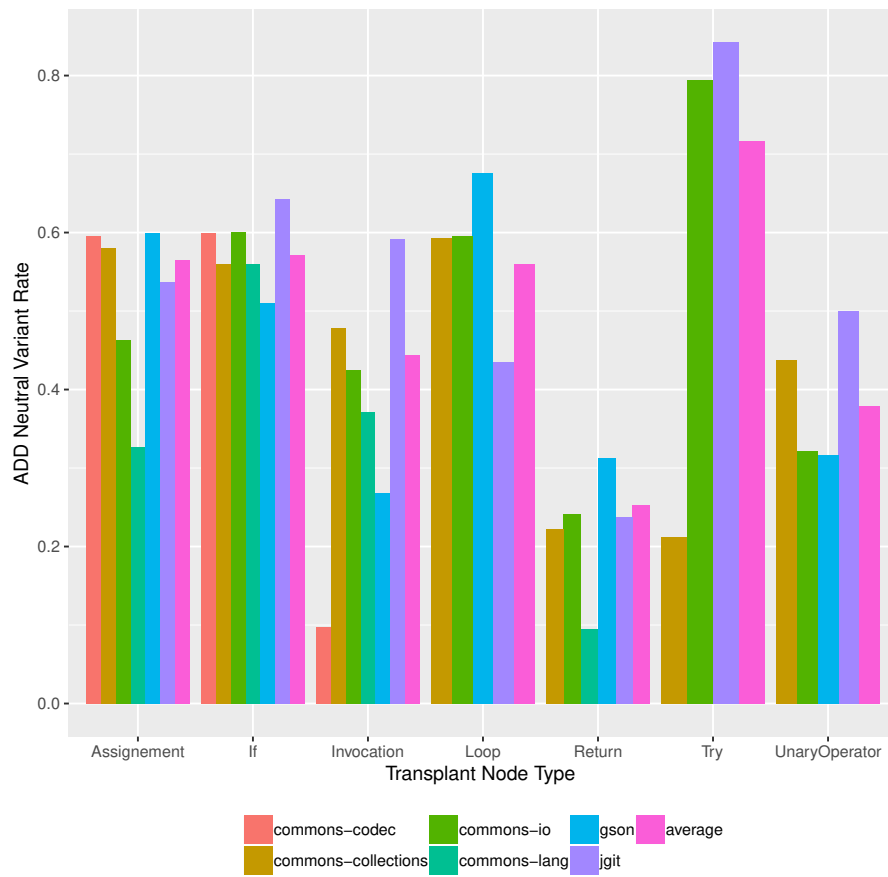}
	\caption{Neutral variant rate for the \add transformation, depending on the type of AST node used as transplant}
	\label{fig:add}
\end{figure}

\autoref{fig:add} displays the neutral variant rate of the \add transformation according  to the type of statements added  (type of the transplant node in the AST). 
Each cluster of bars includes one bar per case study. The darkest bar represents the average neutral variant rate.
\nmodified{The figure only displays the distributions for the node types for which we  performed more than 25 trial transformations for a given project.}

The first striking observation is that neutral variant rates reach significantly high values. In four cases, the random addition of statements yields more than 60\% neutral variants: add  ``if'' nodes in \texttt{jgit}, add ``loop'' nodes in \texttt{gson} and ``try'' nodes in both \texttt{commons-io} and \texttt{jgit}. The addition of such nodes provides important opportunities to explore alternative executions.

We observe important variations between node types as well as between projects. However, some regularities emerge: for instance, adding a ``return'' always yields a low neutral variant rate. This low plasticity of return statements matches the intuition: this is the end point of a computation and it is usually a region where a very specific behavior is expected (and formalized as an assertion in the test).
Meanwhile, the addition of ``Try'' statements appears as an effective strategy to generate neutral variants.

Looking more closely at \autoref{fig:add}, we realize that on average, the addition of ``assignment'' nodes is the most effective (if we exclude addition of ``try'' nodes for which we don't have enough data for all projects). 
This can be explained by the fact that there are many places in the code where the variable declaration and the first value assignment for this variable are separated by a few statements. 
In these situations it is possible to assign any arbitrary value to the variable, which will be canceled by the subsequent assignment. Yao and colleagues observed a similar phenomenon of specific assignments that ``squeezes out'' a corrupted state \cite{yao14}).
Also, for some projects, such as \texttt{commons-io} and \texttt{jgit}, the addition of ``invocation'' nodes is effective.
It probably indicates a non-negligible proportion of side-effect free methods in the program, but further experimentation on that matter is detailed in \autoref{sec:addmi}.

The addition of conditionals and loops is also effective. It is important to understand that a large number of these additional blocks have conditions such that the execution never enters the body of the block, meaning that only the evaluation of the condition is executed.

\subsubsection{\delete}

\begin{figure}[ht]
	\centering
	\includegraphics[width=\columnwidth]{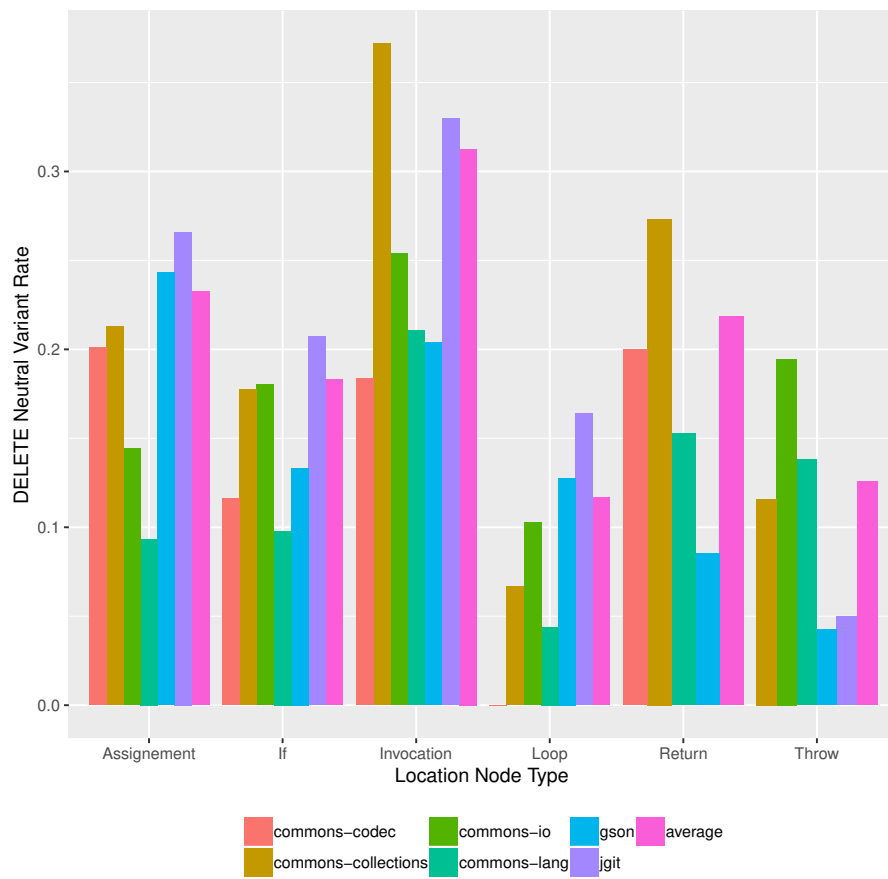}
	\caption{Neutral variant rate of \delete transformation in function of the type of the location}
	\label{fig:delete}
\end{figure}

\autoref{fig:delete} shows the neutral variant rate of the \delete transformation in function of the type of the AST node deleted, grouped by project. The figure only shows the node types for which enough data were collected \nmodified{(More than 25 transformations tried for a given project).}
While we observe large variations between projects for a given node type, we also note that there is a large variation in the neutral variant rate per node type.
For instance, this figure suggests that method invocations are less specified than while-blocks, since the neutral variant rate is higher.

It appears that deleting a method invocation produces above average results for all projects of our sample. We explain this effect by the presence of side-effect free methods which can be safely removed (discussed also in the next section) and by the existence of many redundant calls  (discussed in the next section). 

The deletion of ``continue'' nodes is quite effective at synthesizing neutral variant as it yields 27\% success overall \nmodified{(Not included on the graph since not enough trials were conducted per project, even if over all projects 102 trials were done.). }
Those nodes are usually used as shortcuts in the computation, hence removing them yields slower yet acceptable program variants; we discuss this in depth in the next section. 

\subsubsection{\replace}
\label{sec:replace}

A \replace transformation can be seen as the combination of a \delete and an \add. Consequently, results are somewhat similar to the ones of \add and \delete. 
The neutral variant rate can be seen as the probability that the outcome of a transformation that compiles also passes the tests. This means that if \add and \delete transformations were independent for a given statement, the neutral variant rate for \replace should be close to the product of the two others.
Yet, for each project (as shown in \autoref{tab:adrsr}), the neutral variant rate for \replace is higher than this product, meaning that local neutral variant rate of \add and \delete are probably not independent.

\begin{table}[ht]
  \centering
  \small{
    \begin{tabularx}{\columnwidth}{lXXX}
    \hline
       & \add NVR & \delete NVR & \replace NVR \\ 
    \hline

      commons-codec & $45.51\%\pm3.87$ & $20.03\%\pm2.91$ & $10.55\%\pm1.2$ \\ 
      commons-collections & $53.14\%\pm1.14$ & $23.63\%\pm1.47$ & $13.63\%\pm0.39$ \\ 
      commons-io & $51.74\%\pm1.68$ & $19.35\%\pm1.91$ & $12.53\%\pm0.6$ \\ 
      commons-lang & $42.45\%\pm3.08$ & $12.99\%\pm1.72$ & $11.05\%\pm0.88$ \\ 
      gson & $48.04\%\pm1.45$ & $18.6\%\pm2.24$ & $16.74\%\pm0.69$ \\ 
      jgit & $58.29\%\pm1.68$ & $26.7\%\pm1.21$ & $23.86\%\pm0.75$ \\ 
    \hline
      Total & $51.83\%\pm0.69$ & $22.48\%\pm0.71$ & $15.42\%\pm0.26$ \\ 
     \hline
  \end{tabularx}
  }
  \caption{\modified{Neutral variant rate of \add, \delete, and \replace by project and their 95\% confidence interval}}
  \label{tab:adrsr}
\end{table}

We note two key phenomena.
First, picking a transplant and a location that are method invocations is quite effective. This suggests the presence of alternative yet equivalent calls. This is similar to what is discussed in the next section and also  by Carzaniga \textit{et al}. \cite{Carzaniga14}. 
It also appears that replacing an assignment by another one is efficient. 
Second, we observe a certain plasticity around ``return'' statements: some of them can be replaced by the statement surrounded by a ``try'' or a condition. This suggests the existence of similar statements in the neighborhood of the location, which perform additional checks.

\begin{framed}
	Answer to RQ3: Generic, random \transfs can yield more than $23.30\%\pm0.26$ neutral program variants, but not all code regions are equally prone to neutral variant synthesis. In particular, method invocations and variable assignments are more plastic than the rest of the code.
\end{framed}

\subsection{Role of plastic code regions}
\label{sec:taxonomy}

This section focuses on RQ4. Now, we are interested in understanding whether there is a difference in nature between the neutral variants and the variants that fail the test suite.

\begin{framed}
	RQ4. What roles do the code regions prone to neutral variant synthesis play in the program?
\end{framed}

For each program, we selected neutral variant among extreme cases: those  synthesized on locations covered by a single test case or synthesized on points covered by the highest number of test cases.
By doing this, we are able to build a taxonomy of neutral variants.

This analysis is the result of more than two full weeks of work, where we have manually analyzed dozens of neutral variants. 
At a very coarse grain, before explaining them in detail, we distinguish three kinds of neutral variants:
\begin{inparaenum}[(i)]
  \item \emph{revealer neutral variants} indicate the presence of software plasticity in the code; 
  \item \emph{fooler neutral variants} are named after Cohen's \cite{cohen93} counter-measures for security. 
  \item \emph{buggy neutral variants} are made on locations that are poorly specified by the test suite, the transformation simply introduces a bug.
\end{inparaenum}

\emph{Revealer neutral variants} take their denomination from the fact that they reveal something in the code that is implicit otherwise: \emph{code plasticity}.
Once those regions are revealed,  \transf can target them, with a high confidence that the variant shall be neutral.    

\emph{Fooler neutral variants} are called like this in reference to the ``garbage insertion'' transformation proposed by Cohen \cite{cohen93}. These neutral variants add garbage code that can fool attackers who look for specific instruction sequences. To this extent,  neutral variant synthesis  can be seen as a realization of Cohen's transformation.

\emph{Buggy neutral variants} are simply the degenerated and uninteresting by-products resulting from of weak test cases.  We will not provide a taxonomy of buggy neutral variants. 

In the following, we  discuss categories of revealer and fooler neutral variants. For each category, we present a single archetypal example from the ones synthesized for this work (\autoref{tab:overall-results}). 
Each example illustrates the difference in the original that produces a neutral variant. Examples come with a table that provides the values for the location features. A more complete set of examples is available online. \footnote{\scriptsize \url{https://github.com/castor-software/journey-paper-replication/tree/master/RQ4}}

\textbf{Plastic specification.} Some program regions implement behavior which correctness is not binary. In other terms, there is no one single possible correct value, but rather several ones. We call such specification ``plastic''. 

The regions of code implementing plastic specifications  provide great opportunities for the synthesis of neutral variants, which transform the programs in many ways  while maintaining valuable and correct-enough functionality. 

One situation that we have encountered many times relates to the production of hash keys. 
Methods that produce these keys have a very plastic specification: they must return an integer value that can be used to identify an element. The only contract  is that the function must be deterministic. 
Otherwise, there is no other constraint on the value of the hash key. 
\autoref{lst:coll-hash} illustrates an example of a neutral variant synthesized by removing a statement from a hash method (line \ref{hash-removed}). To us, the neutral variant still provides a perfectly valid functionality.

\begin{figure}[ht]
\begin{lstlisting}[caption={Delete a statement in \texttt{hash} (commons.collection)},label={lst:coll-hash},numbers=left,language=Java,escapechar=@]
  int hash(final Object key) {
    int h = key.hashCode();
@\textcolor{red}{-}@   @\textcolor{red}{h += ~(h << 9);}\label{hash-removed}@
    h ^=  h >>> 14;
    h +=  h << 4;
    h ^=  h >>> 10;
    return h;}
\end{lstlisting}

\tabcolsep=0.11cm
\scriptsize
\begin{tabular}{>{\small}c>{\small}c>{\small}c}
\hline
\rowcolor{lightgray} \#tc &   transfo type & node type   \\
\hline
 422 &  del & var declaration  \\
\hline
\end{tabular}
\end{figure}

\textbf{Optimization} Some code is purely about optimization, which is an ideal plastic region. If one removes such code, the output is still exactly the same, only non-functional properties such as performance are impacted. 
\autoref{lst:range-tostring} shows an example of neutral variant that removes an optimization: at the end of the \texttt{if-block} (line \ref{tostring-removed}), the original program stores the value of \texttt{buf} in  \texttt{toString}, which allows to bypass the  computation of \texttt{buf} next time \texttt{toString()} is called; the neutral variant removes this part of the code, producing a potential performance degradation if the method is called intensively.

\begin{figure}[ht]
\begin{lstlisting}[caption={Delete a statement in \texttt{toString} (commons.lang)},label={lst:range-tostring},numbers=left,escapechar=@]
  String toString() {
    String result = toString;
    if (result == null) {
      final StringBuilder buf = new StringBuilder(32);
      [...] //...compute buf
      result = buf.toString();
@\textcolor{red}{-}@     @\textcolor{red}{toString = result;} \label{tostring-removed}@
    }
    return result;}
\end{lstlisting}
\tabcolsep=0.11cm
\scriptsize
\begin{tabular}{>{\small}c>{\small}c>{\small}c}
\hline
\rowcolor{lightgray} \#tc &   transfo type & node type   \\
\hline
2&   del &stmt list  \\
\hline
\end{tabular}
\end{figure}

\textbf{Code redundancy.} Sometimes, the very same computation is performed several times in the same program. For instance, two subsequent calls to \texttt{list.remove(o)}, even separated by  other instructions are equivalent (as long as \texttt{list} and \texttt{o} do not change between).
\ctransfs naturally exploit this  computation redundancy through the removal or replacement of these redundant statements. 
\modified{Replacement with a call to a side-effect free method also produces valid neutral variants.}

\autoref{lst:center} displays an example of such a neutral variant (removing if-block at line \ref{center-removed}). 
The statement \texttt{if (isEmpty(padStr)) { 
  padStr = SPACE;
} }  assigns a value to  \texttt{padStr}, then this variable is passed to methods \texttt{leftPad} and \texttt{rightPad}. Yet, each of these two methods include the exact same statement, which will eventually assign a value to \texttt{padStr}. So, the statement is redundant and can be removed from the original program, yielding a valid fooler neutral variant.
Compared to neutral variants that remove some optimization, those neutral variants might perform better than the original program.

\begin{figure}[ht]
\begin{lstlisting}[caption={Delete in \texttt{center} (commons.lang)},label={lst:center},numbers=left,language=java,escapechar=@]
  String center(String str, final int size, String padStr) {
    if (str == null || size <= 0) {return str;}
@\textcolor{red}{-}@   @\textcolor{red}{if (isEmpty(padStr)) \{padStr = SPACE;\}}\label{center-removed}@
    [...]
    str = leftPad(str, strLen + pads / 2, padStr);
    str = rightPad(str, size, padStr);
    return str;}
\end{lstlisting}
\tabcolsep=0.11cm
\scriptsize
\begin{tabular}{>{\small}c>{\small}c>{\small}c}
\hline
\rowcolor{lightgray} \#tc &   transfo type & node type   \\
\hline
1&  del & if \\
\hline
\end{tabular}
\end{figure}

\textbf{Implementation redundancy.} It often happens that programs embed several different functions that provide the same service, in different ways. For example, there can exist several versions of the same method with different sets of parameters, which can be used interchangeably by providing good parameter values. It is also possible to use libraries that provide this diversity of similar methods (as demonstrated by Carzaniga and colleagues \cite{Carzaniga14}). 
\autoref{lst:utils-get} illustrates the exploitation of such implementation redundancy inside the program (replace at line \ref{get-replace}), i.e., \texttt{((Object[]) object)[i]} has the same behavior as \texttt{Array.get(object, i)}, with completely different implementations.

\begin{figure}[ht]
\begin{lstlisting}[caption={Replace in \texttt{get} (commons.collection)},label={lst:utils-get},numbers=left,escapechar=@]
  Object get(final Object object, final int index) {
    [...]
    else if (object instanceof Object[]) {
@\textcolor{red}{-}@     @\textcolor{red}{return ((Object[]) object)[i];}\label{get-replace}@
@\textcolor{green}{+}@     @\textcolor{green}{try \{}@
@\textcolor{green}{+}@       @\textcolor{green}{return Array.get(object, i);}@
@\textcolor{green}{+}@     @\textcolor{green}{\} catch (final IllegalArgumentException ex) \{}@
@\textcolor{green}{+}@       @\textcolor{green}{throw new IllegalArgumentException("Unsupported }@
@\textcolor{green}{+}@       @\textcolor{green}{ object type: " + object.getClass().getName());}@
@\textcolor{green}{+}@     @\textcolor{green}{\}}@
    } 
    [...]
  }
\end{lstlisting}
\tabcolsep=0.11cm
\scriptsize
\begin{tabular}{>{\small}c>{\small}c>{\small}c}
\hline
\rowcolor{lightgray} \#tc &   transfo type & node type   \\
\hline
1 &  rep & return \\
\hline
\end{tabular}
\end{figure}

\textbf{Optional functionality.}  
In software, not all parts are of equal importance. Some parts represent the core functionality, other parts are about options and are not essential to the computation. Those optional parts are either not specified or the specification is of less importance. 
These are areas that can be safely removed or replaced while still producing useful variants. 
\autoref{lst:canonicalize} is an example of neutral variant that exploits such optional functionality.
The neutral variant completely removes the body of the method, which is supposed to transform the type passed as parameter into an equivalent version that is serializable, and instead it returns the parameter. 
The neutral variant is covered by 624 different test cases, it is executed 6000 times and all executions complete successfully, and all assertions in the test cases are satisfied. 
This is an example of an advanced feature implemented in the core part of GSon that is not necessary to make the library run correctly.

\begin{figure}[ht]
\begin{lstlisting}[caption={Replace in \texttt{canonicalize} (GSon)},label={lst:canonicalize},numbers=left,escapechar=@]
  public static Type canonicalize(Type type) {
@\textcolor{red}{-}@     @\textcolor{red}{if (type instanceof Class) \{ }@
@\textcolor{red}{-}@       @\textcolor{red}{Class<?> c = (Class<?>) type;}@
@\textcolor{red}{-}@       @\textcolor{red}{return c.isArray() ? new}@
@\textcolor{red}{-}@       @\textcolor{red}{GenericArrayTypeImpl(canonicalize(c.getComponentType())) : c;}@
@\textcolor{red}{-}@     @\textcolor{red}{\}}@
@\textcolor{red}{-}@     @\textcolor{red}{else}@
@\textcolor{red}{-}@     @\textcolor{red}{if (type instanceof ParameterizedType) \{}@
@\textcolor{red}{-}@       @\textcolor{red}{ParameterizedType p = (ParameterizedType) type;}@
@\textcolor{red}{-}@       @\textcolor{red}{return new ParameterizedTypeImpl(p.getOwnerType(),}@
@\textcolor{red}{-}@       @\textcolor{red}{p.getRawType(), p.getActualTypeArguments());}@
@\textcolor{red}{-}@     @\textcolor{red}{\} }@
@\textcolor{red}{-}@     @\textcolor{red}{else }@
@\textcolor{red}{-}@     @\textcolor{red}{if (type instanceof GenericArrayType) \{}@
@\textcolor{red}{-}@       @\textcolor{red}{GenericArrayType g = (GenericArrayType) type;}@
@\textcolor{red}{-}@       @\textcolor{red}{return new GenericArrayTypeImpl(g.getGenericComponentType());}@
@\textcolor{red}{-}@     @\textcolor{red}{\} }@
@\textcolor{red}{-}@     @\textcolor{red}{else }@
@\textcolor{red}{-}@     @\textcolor{red}{if (type instanceof WildcardType) \{}@
@\textcolor{red}{-}@       @\textcolor{red}{WildcardType w = (WildcardType) type;}@
@\textcolor{red}{-}@       @\textcolor{red}{return new WildcardTypeImpl(w.getUpperBounds(),}@   
@\textcolor{red}{-}@       @\textcolor{red}{w.getLowerBounds());}@
@\textcolor{red}{-}@     @\textcolor{red}{\} }@
@\textcolor{red}{-}@     @\textcolor{red}{else \{}@
@\textcolor{red}{-}@       @\textcolor{red}{return type;}@
@\textcolor{red}{-}@     @\textcolor{red}{\} }@
@\textcolor{green}{+}@     @\textcolor{green}{return type; }@
  }
\end{lstlisting}
\tabcolsep=0.11cm
\scriptsize
\begin{tabular}{>{\small}c>{\small}c>{\small}c}
\hline
\rowcolor{lightgray} \#tc &   transfo type & node type   \\
\hline
623 &  rep & if   \\
\hline
\end{tabular}
\end{figure}

\begin{figure}[ht]
\begin{lstlisting}[caption={Add in \texttt{ensureCapacity} (commons.collection)},label={lst:ensureCapacity},numbers=left,escapechar=@]
  void ensureCapacity(final int newCapacity) {
    final int oldCapacity = data.length;
    if (newCapacity <= oldCapacity) {
      return;
    }
    if (size == 0) {
      threshold = calculateThreshold(newCapacity, loadFactor);
      data = new HashEntry[newCapacity];
    } else {
      [...]
    }
@\textcolor{green}{+}@   @\textcolor{green}{ensureCapacity(threshold);}\label{ensure-add}@
  }
\end{lstlisting}
\tabcolsep=0.11cm
\scriptsize
\begin{tabular}{>{\small}c>{\small}c>{\small}c}
\hline
\rowcolor{lightgray} \#tc &   transfo type & node type   \\
\hline
 8&   add & invocation \\
\hline
\end{tabular}
\end{figure}

\newpage
\textbf{Fooler neutral variants.}
We have realized that a number of \add and \replace transformations result in neutral variants which have more code than the original and where the additional code is harmless for the overall execution. 
These neutral variants act exactly as Cohen's ``garbage insertion'' strategy to fool malicious attackers, hence we call them fooler neutral variants.

We found multiple kinds of fooler neutral variants: some add branches in the code or redundant method calls or redundant sequences of method invocations. Some others reduce the legitimate input space through additional checks on input parameters.
\autoref{lst:ensureCapacity} is an example of a fooler neutral variant, which adds a recursive call to \texttt{ensureCapacity()} (line \ref{ensure-add}). 
This could turn the method into an infinite recursion, except that in the additional recursive invocation, the value of the parameter is  such that the condition of the first \texttt{if} statement  always holds true and  the method execution immediately stops.
The additional invocation adds a harmless method call in the execution flow.

\pagebreak
\textbf{Discussion} 
Let us now consider again the location features given for each neutral variant. 
Most neutral variants manually identified as buggy occur on locations covered by a single test case. 
In other words,  the risk of synthesizing bad neutral variants increases when the number of test cases is low.

More interestingly, we realized that valid revealer and fooler neutral variants can be found both on points intensively tested and on weakly tested points.
This confirms the intuition we expressed in the previous section: if a region is intrinsically plastic (has a plastic specification or is optional), the number of test cases barely matters, the only fact that the specification and the corresponding code region is plastic explains the fact that we can easily synthesize neutral variants.

\begin{framed}
	Answer to RQ4: We have provided a first classification of plastic code regions according to the role this region plays in a program. The ``revealers'' indicate plastic code regions \cite{rinard2012}. The ``foolers'' are useful in a protection setting \cite{cohen93}. 
	Our manual analysis shows the variety of roles that code plays in a program. It uncovers the multitude of opportunities that exist to modify the execution of programs while maintaining a global, acceptable functionality.
\end{framed}

\newpage
\subsection{Targeted transformations}
\label{sec:targeted}

\begin{framed}
  RQ5. Can \transfs target specific plastic code regions in order to increase their capacity of synthesizing neutral variants that exhibit behavioral variations?
\end{framed}

For this question we  design three novel, targeted \transfs: \addmi that adds an invocation at the location, \swap that modifies the type of concrete objects that are passed to variables declared with an abstract type, and \loopflip that reverses the order in which a loop iterates over a sequence of elements. These transformations refine the previous \add, \delete, \replace to target language constructs that are most likely plastic regions. Our intention is to design transformations that are more likely to produce variants that are syntactically correct, pass the same test suite as the original and exhibit a behavior that is different from the original. 
We assess the effectiveness of each targeted transformation with respect to:
\begin{itemize}
\item neutral variant rate, as defined in \autoref{def:SR}
\item behavior difference
\end{itemize}

We assess behavior difference by comparing the  traces produced by the original and the neutral variant when running with the same input.
For each targeted transformation, we select the relevant trace features that must be collected, in order to  tune yajta (cf. \autoref {sec:tools}). Then, the traces are aligned up until the first execution of the transformed region. If the traces diverge between a neutral variant and the original, we consider that the \transf has, indeed, yield an observable behavioral difference. 
This reveals that
\begin{inparaenum}[i)]
\item the transformation was performed on code that is not dead; 
\item the compiler optimizations did not mask the effect of the transformations; and
\item two different executions can yield the same result.
\end{inparaenum}
\nmodified{The assessment of behavioral differences through execution traces has proven useful in the search for patches fixing bugs in the field of automatic program repair \cite{deSouza2018}}

\subsubsection{\addmi}
\label{sec:addmi}

The \addmi transformation leverages the following  observation:  \autoref{fig:add} indicates that the addition of ``invocation'' nodes is likely to produce neutral variants. We focus on invocations rather than loops or conditions to reduce the risk of synthesizing variants where the added code is not executed.
We also exploit the good results obtained when adding ``try'' blocks.

\subsubsection*{The \addmi transformation process}

The transformation starts with the  selection of a random location $\pi$. 
Then, it builds the set of  methods that are accessible from $\pi$. A method is considered to be accessible if
\begin{inparaenum}[(i)]
	\item  the method is public, protected and in the same package as the class of $\pi$, or private and in the same class; 
	\item if $\pi$ belongs to the body of a static method, the method called must be static.
    \item if $\pi$ does not belong to the body of a static method, the inserted invocation must either refer to a static method, or refer to a method member of the class of an object available in the context.
	\item there exists a set of variables  in the local context to fit the method's parameters.	
\end{inparaenum}
Let us notice that we prevent the method hosting $\pi$ to be selected, as this would create recursive calls likely to produce an infinite loop.

Once a method $m$ has been selected, we synthesize a transplant in the form of an invocation AST node to insert at the location. If the return type of $m$ is not void, a public field is synthesized in the hosting class and the invocation result is assigned to this field. This additional rule aims at forcing the usage of the invocation's result and hence at preventing the compiler from considering the invocation as dead code and removing it \cite{rodriguezcancio:hal-01343818}. 
The transplant is then wrapped into a ``try-catch'' block.

\modified{A formal definition of the transformation is provided in \autoref{sec:appendix}.}

\subsubsection*{Illustration of the \addmi transformation}

\autoref{lst:ex_addMI} illustrates the addition of an invocation of  \texttt{conditionC0(String, int)}  before the \texttt{return} statement. Since \texttt{conditionC0} returns a boolean, a public field of the same type is added to the  \texttt{DoubleMetaphone} class to consume the result of the invocation.

\begin{lstlisting}[caption={Add method invocation in \texttt{DoubleMetaphone.java:882} (commons.codec)},label={lst:ex_addMI},numbers=left,language=Java,escapechar=@,firstnumber=882]
@\textcolor{green}{+}@ @\textcolor{green}{public boolean v6482819 = true;}@
  private boolean isSilentStart(final String value) {
    boolean result = false;
    for (final String element : SILENT_START) {
      if (value.startsWith(element)) {
        result = true;
        break;
      }
    }
@\textcolor{green}{+}@   @\textcolor{green}{try \{}@
@\textcolor{green}{+}@     @\textcolor{green}{v6482819 =  conditionC0(this.VOWELS, this.maxCodeLen);}@
@\textcolor{green}{+}@   @\textcolor{green}{\} catch (Exception v4663426) \{\}}@
    return result;
  }
\end{lstlisting}

\begin{figure}[ht]
  \centering
  \includegraphics[width=0.9\columnwidth]{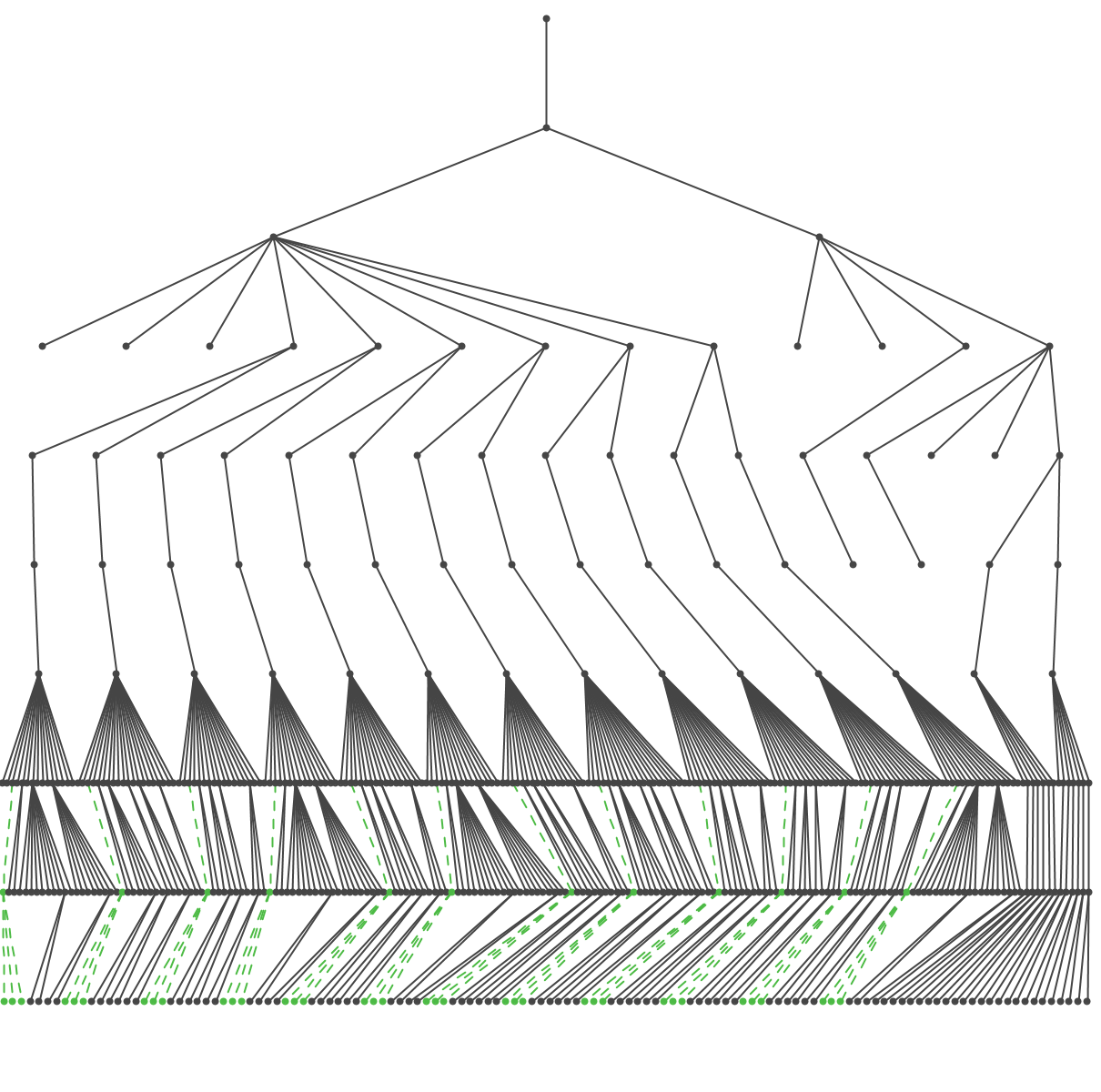}
  \caption{Impact of a modification on StringEncoderComparatorTest call tree}
  \label{fig:addMI_tree}
\end{figure}

\autoref{fig:addMI_tree} illustrates the juxtaposition of two dynamic call trees: the tree of the execution of \texttt{StringEncoderComparatorTest} on the original \texttt{isSilentStart} method and the tree when running the same test on the transformed method. Each node on the figure represents  a method  and each edge represents a method invocation. The temporal aspect of the execution is represented in two dimensions:  method invocations go from top to bottom, and, if a method invokes several others, the calls on the left occur before those on the right.
The nodes in grey represent calls the parts of the test execution that are common to both the original and the transformed program. Nodes in light green (and connected with dashed lines) represent the parts of the execution added with the transformation.

\subsubsection*{Searching the space of the \addmi transformation} 

The size of the search space can be bound by the product of the number of statements in the targeted program and the number of methods it declares. In practice, we limit ourselves to methods for which we can pass parameters within the context of the location, which significantly reduces the size of the space. Yet, the space remains huge. Consequently, for experimental purposes, we limit our search to up to 10 different methods per location. If more than 10 methods can be invoked at the same point, we randomly select 10. \modified{As we performed sampling on the search space, in the rest of the subsection, we give NVR measures with their 95\% confidence interval modelled following a binomial distribution.}

\newpage
\subsubsection*{Behavior diversity}

To assess the behavioral variations introduced by the addition of a method invocation, we use \textbf{yajta} to trace the number of times each method in the program invokes any other method. This observation produces a $N \times N$ matrix, where N is the number of methods executed when running the test suite. The comparison of the matrix produced on the original program and the one produced on the variant reveals if it is, indeed, possible to observe additional method invocations (i.e., additional behavior) at runtime.

\begin{table}[ht]
	\caption{Call Matrix of the execution of \texttt{StringEncoderComparatorTest} when adding a call to \texttt{conditionC0} in \texttt{isSilentStart} (in DoubleMetaphone)}
    \begin{tabular}{r|ccccccc}
      &
      \rotatebox{-90}{   ...    } &
      \rotatebox{-90}{isSilentStart(String)} &
      \rotatebox{-90}{conditionC0(String, int)} &
      \rotatebox{-90}{contains(String, int, int, String[])} &
      \rotatebox{-90}{charAt(String, int))} &
      \rotatebox{-90}{isVowel(char)} &
      \rotatebox{-90}{   ...    } \\
    \hline
      ...            & ...   & ...   & ...   &  ...   &  ...   &  ...   &  ...   \\ 
      isSilentStart(String)        & ...   & 0     & 0 \textcolor{green}{+12}     &  0    &  0   &  0   &  ...   \\ 
      conditionC0(String, int)       & ...   & 0     & 0     &  0 \textcolor{green}{+12} &  0 \textcolor{green}{+12} &  0 \textcolor{green}{+12}     &  ...   \\
      doubleMetaphone(String, boolean)        & ...   & 12 \textcolor{green}{+0}     & 0   & 0   & 2 \textcolor{green}{+0}    &  0    &  ...   \\
      ...             & ...   & ...   & ...   &  ...   &  ...   &  ...   &  ...   \\
    \hline
    \end{tabular}
	\label{tab:addmi-matrix}
\end{table}

\autoref{tab:addmi-matrix} shows an excerpt of  the  trace when running  \texttt{StringEncoderComparatorTest}. Each line records the number of times a method has invoked the methods mentioned in the column header. The results recorded during the execution of the test on the original program appear in black, while the new calls, occurring as a result of the transformation, appear in green. We observe that the transformed method (\texttt{isSilentStart}) is called 12 times by \texttt{doubleMetaphone} during the test run, on the original program. The \transf adds an invocation to \texttt{conditionC0} in  \texttt{isSilentStart}. This results in 12 invocations of \texttt{conditionC0}, as well as 12 times more invocations to all the methods invoked by \texttt{conditionC0}.  These can be observed in \autoref{fig:addMI_tree} as 12 subtrees of one node calling 3 other appear in green.

\newpage
\subsubsection*{Empirical results for the \addmi transformation}

\newcommand{\nbnvaddmi}{171 744\xspace}
\newcommand{\nbcompileaddmi}{259 099\xspace}
\newcommand{\nvraddmi}{66.29\%\xspace}
\modified{
\autoref{tab:addmi-results} displays the results per study object: (\#Locs) number of locations for which transformations were attempted, number of times we performed the \addmi transformation and produced a compilable variant (\# Compile); number of transformations that yield a neutral variants (\# NV); and the neutral variant rate (NVR). Overall, \nvraddmi of the \transfs yield a program variant that compiles and passes the suite, which corresponds to \nbnvaddmi neutral variants in total.
}

\begin{table}[ht]
	\centering
	\caption{\modified{Neutral variant rate of \addmi}}
	\small{
\begin{tabularx}{\columnwidth}{lXXXX}
  \hline
    & \#Locs & \#Compiles & \#NV & NVR \\
  \hline
     commons-codec & 1722 & 17650 & 11150 & $63.17\%\pm0.71$ \\ 
     commons-collections & 7027 & 40333 & 26150 & $64.84\%\pm0.47$ \\ 
     commons-io & 1608 & 10009 & 7413 & $74.06\%\pm0.86$ \\ 
     commons-lang & 4287 & 129593 & 86452 & $66.71\%\pm0.26$ \\ 
     gson & 2460 & 32932 & 18215 & $55.31\%\pm0.54$ \\ 
     jgit & 12822 & 28582 & 22364 & $78.25\%\pm0.48$ \\ 
   \hline
     total & 29926 & 259099 & 171744 & $66.29\%\pm0.18$ \\ 
   \hline
\end{tabularx}
	}
	\label{tab:addmi-results}
\end{table}

The first key observation is that method invocations are plastic regions, regardless of the original program. 
 The second observation is that the targeted \transf is significantly \nmodified{(p-value $< 0.001$ with a \texttt{Wilcoxon rank sum test})} more effective than a random invocation addition to synthesize neutral variants: \nvraddmi on average instead of the 45\% neutral variant rate of the \add transformation presented in \autoref{fig:add} when inserting method invocation.
 
 Several factors contribute to this successful synthesis of neutral variants. First, the transformation selects the methods to be added, ensuring that it is possible to get valid parameter values in the context of the location. This design decision can  favor repeating an invocation that already exists in the method that hosts the location. If the method is idempotent, the trace changes with no side effect. Second, the additional invocation is wrapped  into ``try'' blocks. This may also lead to the compilation of invocations that quickly throw an exception and therefore, do not cause any state change.
In general, the addition of invocations to idempotent (i.e. methods that have  no additional effect if they are called more than once with the same input parameters) or pure methods (i.e. method with no externally observable side effect\cite{salcianu2005}) can make the insertion benign.

\begin{table}[ht]
	\centering

	\caption{\modified{Neutral variant rate of \addmi depending on Transplant and Location type}}
	\small{
	\begin{tabularx}{\columnwidth}{l|cc|c}
		\hline
		& Internal Transplant & External Transplant & Total \\
		\hline
			Static Loc & $85.92\%\pm0.62$ & $87.96\%\pm0.52$
 & $87.06\%\pm0.40$ \\
			Non static Loc & $62.52\%\pm0.20$ & $80.89\%\pm0.59$
 & $63.85\%\pm0.20$ \\
		\hline
			Total & $63.75\%\pm0.20$ & $84.25\%\pm0.40$ & $66.29\%\pm0.18$ \\
		\hline
	\end{tabularx}
}
	\label{tab:addmi-results-2}
\end{table}

\newcommand{\nvraddmisi}{85.92\%\xspace}
\newcommand{\nvraddmise}{87.96\%\%\xspace}

\newcommand{\nvraddmini}{62.52\%\xspace}
\newcommand{\nvraddmine}{80.89\%\xspace}

\newcommand{\nvraddmis}{87.06\%\xspace}
\newcommand{\nvraddmin}{63.85\%\xspace}

\newcommand{\nvraddmii}{63.75\%\xspace}
\newcommand{\nvraddmie}{84.25\%\xspace}

In \autoref{tab:addmi-results-2} we provide the cumulative neutral variant rates, with respect to the type of method in which  the location is selected (location (Loc) in static or non-static method) and with respect to the type of transplant (invoke a method that inside the same class as the location or that is external to that class). 
\nmodified{In this table, we observe a significant (p-value $ < 0.001$ with a \texttt{Wilcoxon rank sum test}) difference between the two types of locations: locations in static methods are more plastic (\nvraddmis) that in non static ones (\nvraddmin).} We hypothesize that this comes from the fact that in the case of a location inside a static method, the additional invocation can only be towards a static method. Increased neutral variant rate in this case could come from the fact the proportion of pure methods is higher among static methods than among regular methods.

\nmodified{We also observe more successful transformations when the transplant is selected outside the class that hosts the location (\nvraddmie instead  of \nvraddmii, p-value $< 0.001$ with a \texttt{Wilcoxon rank sum test}).} We hypothesize that methods invoked in the same class as the location are  likely to be non-pure methods.
The transformation selects invocations to methods for which the context of the location can provide values to pass as parameters. This means that most of the methods inside the same class can be invoked, whereas in the case of external methods this tends to select methods with no parameter or methods that have only parameters of primitive data types.
We hypothesize that this difference in the selection of candidate methods increases the chance to have more pure methods among external than among internal method invocations.

\subsubsection{\swap}

The results of the \replace transformation showed that targeting assignment statements yields more neutral variants than on other types of AST nodes. In this 	section, we introduce a new transformation that refines \replace on ``Assignment'', leveraging Java interfaces. A common practice in Java consists of declaring a variable typed with an interface. When a developer adopts this practice, she indicates that any  concrete object that implements the interface can be  assigned to this variable. The existing diversity of available types sharing an interface can be leveraged to fuel our search for neutral variants.

\subsubsection*{The \swap transformation process} 

This \transf operates on assignment statements that pass a new concrete object to a variable typed with an interface. The transformation replaces the constructor called in such assignments by one of a class implementing the same interface. In the following experiments we have implemented this transformation for classes and interfaces of Java collections.

\subsubsection*{Illustration of the \swap transformation}

\begin{figure}[ht]
	\centering
	\includegraphics[width=0.9\columnwidth]{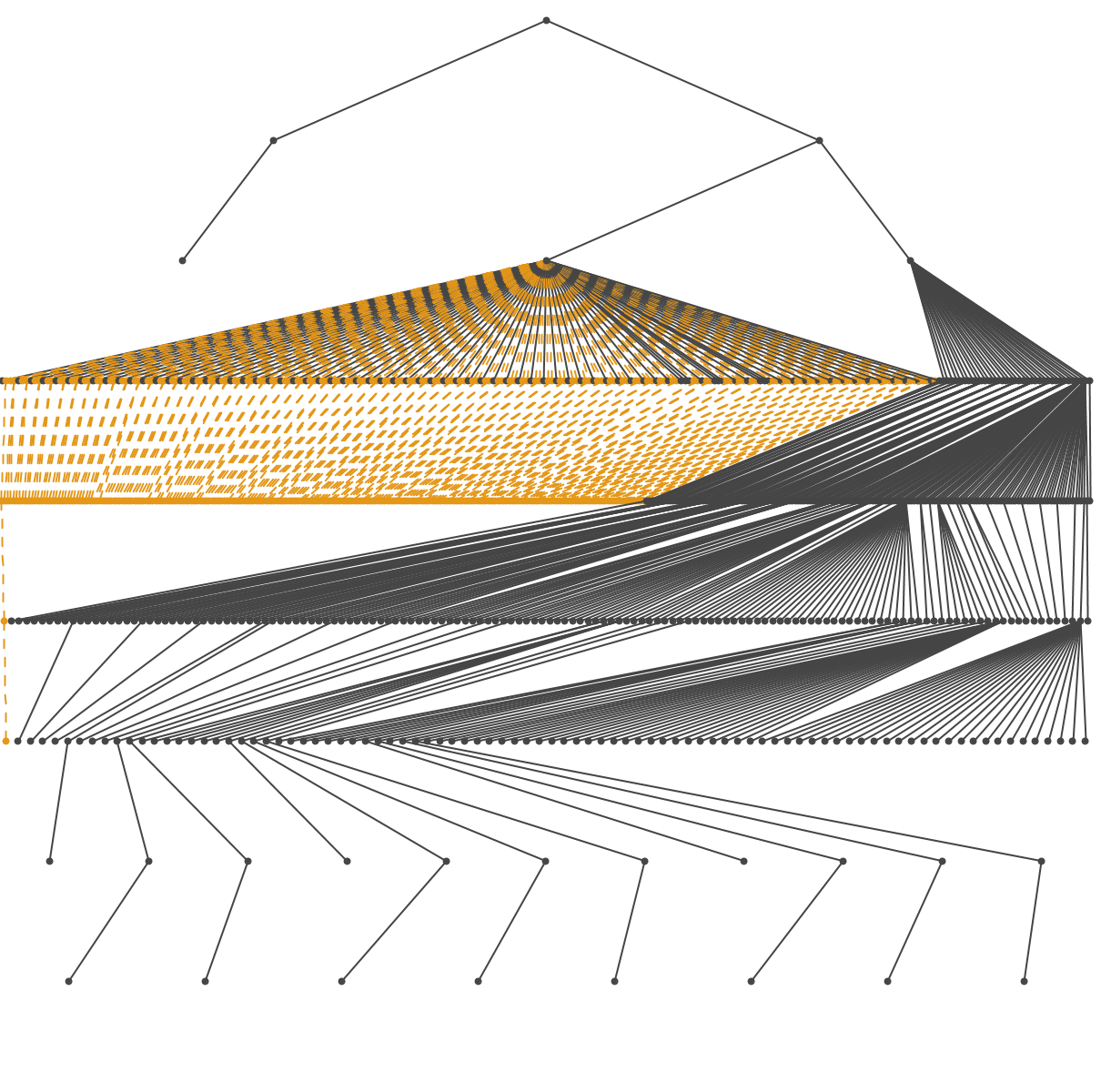}
	\caption{Impact of a modification on the call tree of one execution of \texttt{PhoneticEngineTest.testEncode()}}
	\label{fig:swap_tree}
\end{figure}

\autoref{lst:ex_swap} shows an example of a \swap transformation, while \autoref{fig:swap_tree} illustrates its impact on the dynamic call tree of one test. Nodes in light teal are method invocations  from $org.apache.commons.collections4$ which were not present before. (They replace previous calls to the Java standard library).

\begin{lstlisting}[caption={SwapSubType in \texttt{Lang.java:130} (commons.codec)},label={lst:ex_swap},numbers=left,language=Java,escapechar=@,firstnumber=130]
  public static Lang loadFromResource(final String languageRulesResourceName, final Languages languages) {
@\textcolor{red}{-}@   @\textcolor{red}{final List<LangRule> rules = new ArrayList<LangRule>();}@
@\textcolor{green}{+}@   @\textcolor{green}{final List<LangRule> rules = new org.apache.commons.collections4.list.NodeCachingLinkedList<LangRule>();}@
     [...]
  }
\end{lstlisting}

\subsubsection*{Searching the space of \swap transformations} 

\modified{The search space here is composed of all statements that assign a new concrete object to a variable which type is a collection (see Appendix \ref{sec:swap-appendix} for the actual list).}
This space is small enough to be explored exhaustively. 
We target 16 interfaces, which are implemented by 50 classes (some of which implement several interfaces) from 3 different libraries (\texttt{java.util}, \texttt{org.apache.commons.collections} and \texttt{net.sf.trove4j}).
The complete list of interfaces, and their concrete classes, targeted by this transformation is available in the replication repository. \modified{This transformation is similar to what Manotas and colleagues have implemented in their framework SEEDS\cite{Manotas}.}
While  the choice of a concrete collection might be a long planned decision for performance reasons, we believe that in many cases the choice is  made by default.

\subsubsection*{Behavior diversity}

To observe the changes introduced by the \swap transformation,  we use \textbf{yajta} to trace both the methods defined in the classes of the program that is transformed and  all the methods in collection classes that are involved in the transformation (the ones at the location and the ones in the transplants).
The trace comparison procedure is the same as for the \addmi transformation.

\begin{table}[ht]
	\centering
	\caption{\modified{Neutral variant rate of \swap}}
	\small{
		\begin{tabularx}{\columnwidth}{lXXXX}
			\hline
			& \#Loc & \#Compile & \#NV & NVR \\
			\hline
			    commons-codec & 21 & 186 & 164 & $88.17\%$ \\ 
			    commons-collections & 68 & 738 & 450 & $60.98\%$ \\ 
			    commons-io & 16 & 183 & 177 & $96.72\%$ \\ 
			    commons-lang & 41 & 544 & 445 & $81.80\%$ \\ 
			    gson & 17 & 266 & 221 & $83.08\%$ \\ 
			    jgit & 190 & 2992 & 1403 & $46.89\%$ \\ 
			\hline
			    Total & 339 & 4909 & 2860 & $58.26\%$ \\ 
			\hline
		\end{tabularx}
	}
	\label{tab:swap-results}
\end{table}

\subsubsection*{Empirical results for the \swap transformation} 

\autoref{tab:swap-results} presents the results of the  \swap transformation on each project of our sample. 
In total, we synthesized 4909 variants that compiled on 339 different locations (i.e. collection assignment to a variable typed as an interface for which at least one transformation yields a variant that compiles). Out of the 4909 variants that compile correctly, 2860 are neutral variants. This represents a global 58.26\% neutral variant rate. We notice that the \swap transformation yields more than 80\% neutral variants for 4 projects. Yet, for \texttt{jgit} and \texttt{commons-collections}, the neutral variant rate falls to 46.89\% and 60.98\% respectively. Overall this represents a geometric mean of 74\%. 

A major reason for the lower neutral variant rate  on \texttt{commons-collections} is the use of inner classes that implement the \texttt{Collection} interface. This happens to create classes that mix the contract of the \texttt{Collection} interface with the contract of the class inside which the \texttt{Collection} interface implementation is defined. For example, the class \texttt{MultiValueMap\$Values} implements the iterator of the \texttt{Collection} interface inside \texttt{MultiValueMap}. 
\autoref{lst:ex_swap_failed} shows an instantiation of \texttt{MultiValueMap\$Values} that was used as a location for the \swap transformation. The original program assigns  a \texttt{MultiValueMap\$Values} to  \texttt{valuesView}. This means that subsequent calls to \texttt{MultiValueMap\$Values.iterator()} return the values that are stored in the field \texttt{map}. Now, since \texttt{vs} is a of type \texttt{Collection}, the \swap transformation assumes that it can assign it any object typed with an implementation of \texttt{Collection}, e.g. \texttt{LinkedList} in this example. Yet, because a call to \texttt{iterator()} on an instance of \texttt{LinkedList} only iterate over elements that have been added to the instance, all \texttt{MultiValueMap\$Values.iterator()} calls return empty iterators which leads to failing tests. Such situations occurred for  113 variants, 0 of which are neutral.

\begin{lstlisting}[caption={SwapSubType in \texttt{MultiValueMap.java:326} (commons.codec)},label={lst:ex_swap_failed},numbers=left,language=Java,escapechar=¤,firstnumber=326]
    @Override
    @SuppressWarnings("unchecked")
    public Collection<Object> values() {
        final Collection<V> vs = valuesView;
¤\textcolor{red}{-}¤       ¤\textcolor{red}{return (Collection<Object>) (vs != null ? vs : (valuesView = new Values()));}¤
¤\textcolor{green}{+}¤       ¤\textcolor{green}{return (Collection<Object>) (vs != null ? vs : (valuesView = new LinkedList<V>()));}¤
    }
\end{lstlisting}

While the number of candidates to be targeted by this transformation is lower than for other transformations, \swap affects all subsequent invocations that target the modified variable. Therefore, the \transf impacts the generated variant in a more profound way than other transformations. This effect is well illustrated by \autoref{fig:swap_tree}.

In theory, it is possible to swap any valid subtype of an interface when assigning a concrete object to a variable typed with the interface, and this with no effect on the functionality. This property is a direct consequence of the fact that any requirement on the type of a variable should be expressed in the interface. In other words, \swap should be a sound preserving transformation. 
Indeed, we observe that there exist at least one neutral variant for 71\% of the 339 locations targeted by the \swap transformation.
However, in practice we observe that is not always the case, and \swap is, indeed, a \transf: only 58\% of the transformations actually yield a neutral  variant.

\begin{figure}[ht]
	\begin{lstlisting}[caption={Altering an \texttt{ordered loop} in ReflectiveTypeAdapterFactory:140 (gson)},label={lst:orderedMap},numbers=left,escapechar=@,firstnumber=140]
    List<String> fieldNames = getFieldNames(field);
    BoundField previous = null;
@\textcolor{red}{-}@   @\textcolor{red}{Map<String, BoundField> result = new LinkedHashMap<String, BoundField>();}@
@\textcolor{green}{+}@   @\textcolor{green}{Map<String, BoundField> result = new HashMap<String, BoundField>();}@
    [...]
     for (int i = 0; i < fieldNames.size(); ++i) {
      String name = fieldNames.get(i);
      if (i != 0) serialize = false; // only serialize the default name
      BoundField boundField = createBoundField(context, field, name,
      TypeToken.get(fieldType), serialize, deserialize);
      BoundField replaced = result.put(name, boundField);
      if (previous == null) previous = replaced;
    }
	\end{lstlisting}
\end{figure}

\autoref{lst:orderedMap} illustrates an example where the  \swap transformation fails at producing a neutral variant. Here, the concrete type in the original program is \texttt{LinkedHashMap}. This specific implementation of the \texttt{Map} interface keeps the entries in the order of insertion. When the \texttt{for} loop iterates through the \texttt{fieldNames} list, the \texttt{result} map is filled such that the elements in \texttt{map} are stored in the same order as the elements in \texttt{fieldNames}. Now, when the  \emph{swap subtypes}  transformation  assigns a \texttt{HashMap} object to \texttt{result} instead of a \texttt{LinkedHashMap}, the elements of \texttt{result} are ordered with respect to their hash value instead of keeping the order of insertion of \texttt{fieldNames}. Consequently, subsequent methods that expect a specific order in  \texttt{result} fail because of this change. 

It is important to notice that when we replace  \texttt{LinkedHashMap} by  \texttt{org.apache.} \texttt{commons.collections4.map.LinkedMap} in \autoref{lst:orderedMap}, the corresponding variant is neutral, since the substitute type satisfies the required invariant: elements are kept in order of insertion. More generally, we can say that this zone is plastic, modulo this type invariant.

\subsubsection{\loopflip}

Swapping instructions is a state of the art transformation used by Schulte and colleagues \cite{Schulte13} or in a sound way for obfuscation \cite{wang2017composite}. Here we explore a targeted swap transformation, which reverses the order of iterations in loops.

\subsubsection*{ The \loopflip transformation process} 

We propose a \transf that reverses the order in which  \texttt{for} loops  iterate over a set of elements.
It targets counted loops, i.e., loops for which  we can identify a loop counter variable that is initialized with a specific value and which is increased or decreased at each iteration until it satisfies a condition. 
The transformation does not necessarily expect a well-behaved counted loop. The transformation makes the loop run the same iterations as the original loop, but for loop index values in reverse order.
To achieve this, we need to identify the initial value, the step, and the last value. \autoref{lst:ex_loopflip1} shows such an example for a simple case. The loop counter is the variable $i$, its initial value is $0$, the step is $+ 1$, so the last value is straightforward to determine ($srcArgs.length-1$). In this example, the transformation replaces the original loop with one starting from the last value, with a step of $- 1$ and ending when the loop counter reaches the initial value.

The example of \autoref{lst:ex_loopflip2} is a non-normalized loop that we still handle with \loopflip. The variable $i$ is still the loop counter. Its first value is 0, and its last value is 28 as 32 is not reachable. For the general case, the last value is the last multiple of $step$ smaller than the difference between the upper bound and the starting value. Yet, as we only transform the code in a static way, this expression is directly inserted in the initialization of the loop counter. More implementation details are given in the replication repository.\footnote{\url{https://github.com/castor-software/journey-paper-replication/tree/master/RQ5}}

\begin{lstlisting}[caption={Loopflip in \texttt{MemberUtils.java:115} (commons.lang)},label={lst:ex_loopflip2},numbers=left,language=Java,escapechar=@,firstnumber=115]
@\textcolor{red}{-}@   @\textcolor{red}{for (int i = 0; i < srcArgs.length; i++) \{}@
@\textcolor{green}{+}@   @\textcolor{green}{for (int i = srcArgs.length-1; i >= 0; i--) \{}@
\end{lstlisting}

\begin{lstlisting}[caption={Loopflip in \texttt{UnixCrypt.java:87} (commons.codec)},label={lst:ex_loopflip1},numbers=left,language=Java,escapechar=@,firstnumber=87]
@\textcolor{red}{-}@   @\textcolor{red}{for (int i = 0; i < 32; i += 4) \{}@
@\textcolor{green}{+}@   @\textcolor{green}{for (int i = 32 - (((32 - 0) \% 4) == 0 ? 4 : (32 - 0) \% 4); i >= 0; i -= 4) \{}@
\end{lstlisting}

\subsubsection*{Illustration of the \loopflip transformation}

In order to illustrate how test execution may be affected by this transformation, \autoref{lst:ex_loopflip2} details a transformation, a test that cover the transformed code and its execution trace.

\definecolor{darkgreen}{rgb}{0,0.7,0}
\begin{lstlisting}[caption={Loopflip in \texttt{BinaryCodec.java:108,123} (commons.lang)},label={lst:ex_loopflip_transf},numbers=left,language=Java,escapechar=@,firstnumber=108]
public static byte[] toAsciiBytes(final byte[] raw) {
    if (isEmpty(raw)) {
        return EMPTY_BYTE_ARRAY;
    }
    final byte[] l_ascii = new byte[(raw.length) << 3];
    for (int ii = 0, jj = (l_ascii.length) - 1 ; ii < (raw.length) ; ii++ , jj -= 8) {
@\textcolor{red}{-}@   @\textcolor{red}{for (int bits = 0 ; bits < (BITS.length) ; ++bits) \{}@
@\textcolor{green}{+}@   @\textcolor{green}{for (int bits = (BITS.length - 1) ; bits >= 0 ; --bits) \{}@
            if (((raw[ii]) & (BITS[bits])) == 0) {
@\textcolor{orange}{A}@                @\textcolor{orange}{l\_ascii[(jj - bits)] = '0';}@
            } else {
@\textcolor{darkgreen}{B}@                @\textcolor{darkgreen}{l\_ascii[(jj - bits)] = '1';}@
            }
        }
    }
    return l_ascii;
}
\end{lstlisting}

\autoref{lst:ex_loopflip_transf} shows an example where the \loopflip transformation yields a neutral variants of the  \texttt{BinaryCodec.toAsciiBytes()} method. This method returns an array of \texttt{byte}s, which is filled inside the loop that we transform. The variant is neutral because the array is filled with values which do not depend on the iteration order of the loop, but only on the value of the loop index. We can think of this as filling the table with a set of numbered operations, where the order in which the operations are performed does not matter. Consequently the  \texttt{l\_ascii} array contains exactly the same thing, whatever the iteration order of the \texttt{for} loop in line 114. 

\begin{lstlisting}[caption={Call to toAsciiBytes in \texttt{BinaryCodecTest.java:706,709} (commons.codec)},label={lst:ex_loopflip_running},numbers=left,language=Java,escapechar=@,firstnumber=706]
    bits = new byte[1];
    bits[0] = BIT_0; //0b00010111
    l_encoded = new String(BinaryCodec.toAsciiBytes(bits));
    assertEquals("00010111", l_encoded);
\end{lstlisting}

\begin{figure}[ht]
 \begin{minipage}[b]{0.5\linewidth}
 			\includegraphics[width=\columnwidth]{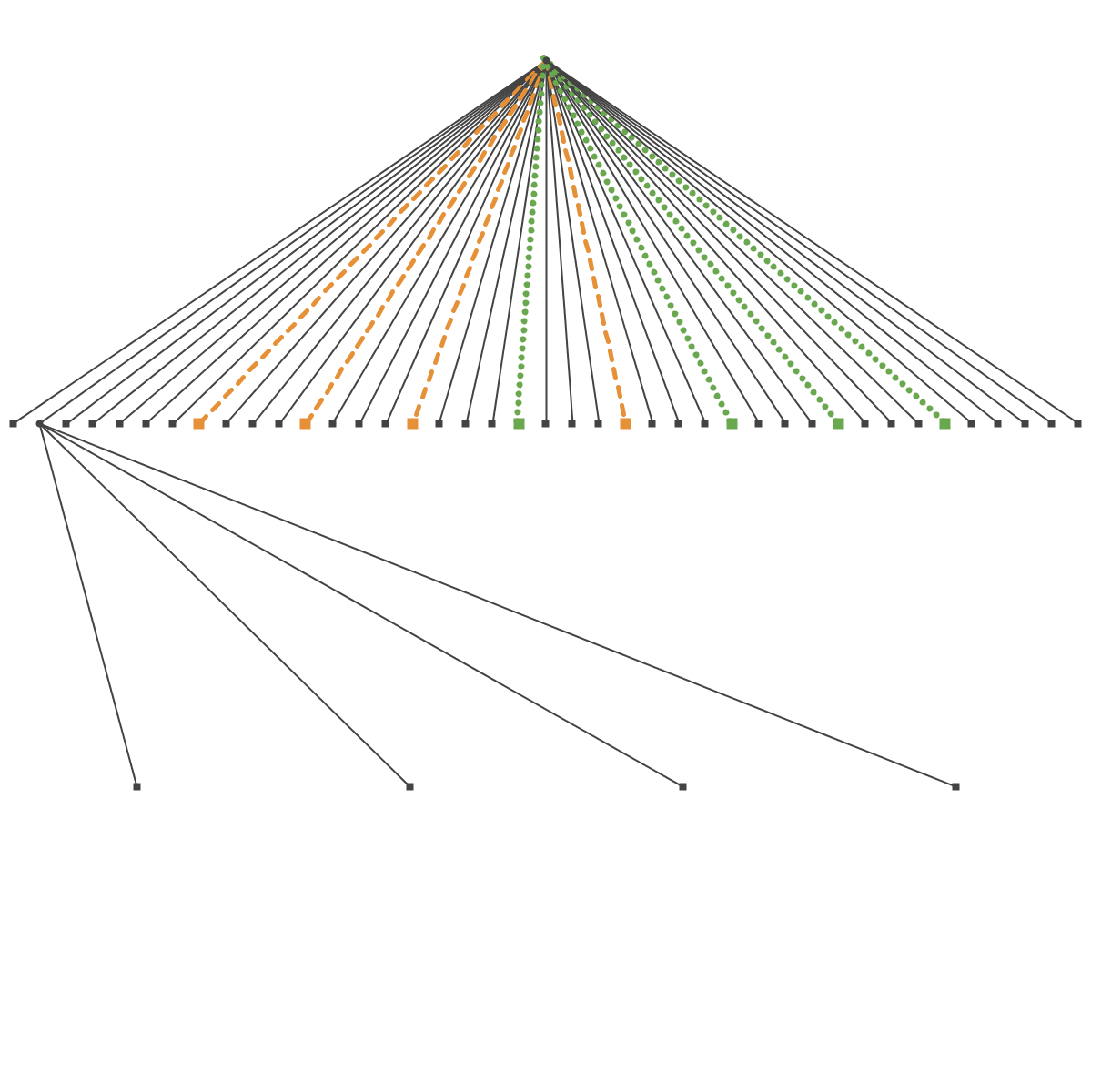}
 			\caption{Original test execution of BinaryCodec.toAsciiBytes. (Note the pattern AAABABBB)}
			\label{fig:loopflipC_tree}

 \end{minipage}
 \hspace{0.5cm}
 \begin{minipage}[b]{0.5\linewidth}
 \centering
 			\includegraphics[width=\columnwidth]{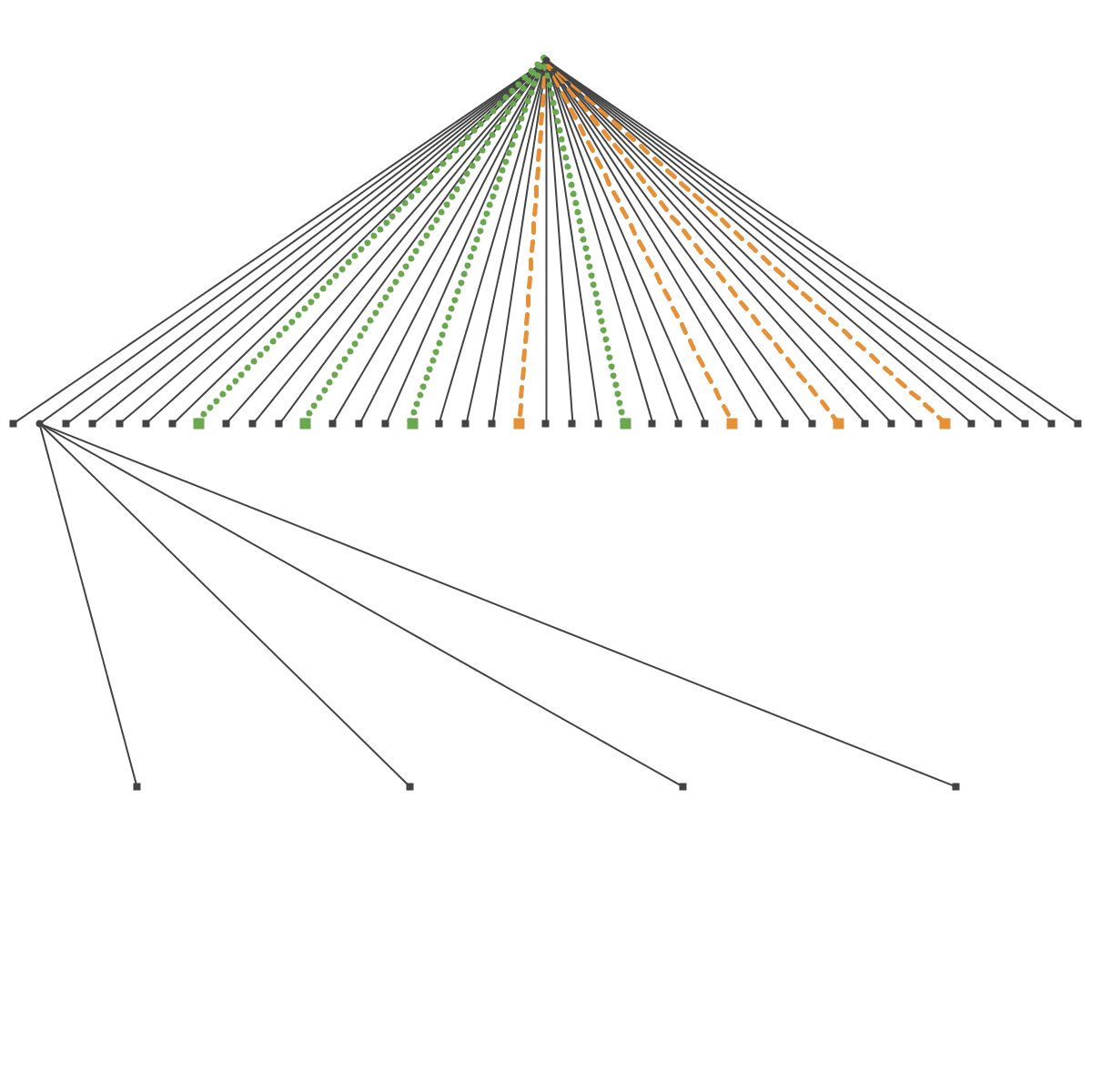}
 			\caption{Transformed test execution of BinaryCodec.toAsciiBytes. (Note the pattern BBBABAAA)}
 			\label{fig:loopflipD_tree}

 \end{minipage}
\modified{Round nodes correspond to method calls, squared ones correspond to branches in the order where they are called (from left to right). The branch highlighted in orange (dashed line) corresponds to the line A (in the same color in \autoref{lst:ex_loopflip_transf}). The branch highlighted in green (dotted line) corresponds to the line B.}
 \end{figure}

\autoref{lst:ex_loopflip_running} shows an excerpt of the test case that specifies the behavior of \texttt{BinaryCodec.toAsciiBytes()}. It calls the method with the binary value $00010111$ as parameter and assesses that the return value is  the  array of \texttt{byte}s that encodes the  String ``00010111''. \autoref{fig:loopflipC_tree} and \autoref{fig:loopflipD_tree} show the execution of both the original and transformed method in that context. Round nodes correspond to method calls, squared ones correspond to branches in  the order where they are called (from left to right). The branch highlighted in orange (dashed line) corresponds to the line A (in the same color in \autoref{lst:ex_loopflip_transf}). The branch highlighted in green (dotted line) corresponds to the line B. We can observe that the execution order is indeed reversed.

\subsubsection*{Searching the space of \loopflip transformations}

In the case of this transformation, the search space is composed of for-loops based on an integer index. Since it is fairly small, we exhaustively explore it.

\subsubsection*{Behavior diversity}

The observation of branch executions would not be enough to systematically detect behavioral differences caused by this transformation for every case. Indeed for a loop whose body is composed of a single branch, branches executed do not depend on the index variable, therefore, branch observation would fail to detect differences. Thus the simplest observation method is to insert a probe at the beginning of the transformed loop to trace the value of the loop index.

\subsubsection*{Empirical results for the \loopflip transformation}

\autoref{tab:loop-flip-results} summarizes the results of the \loopflip experiments. We observe that this \transf is very effective at synthesizing neutral variants. In total, we synthesized 479 neutral variants out of 656 variants, that compiled, targeting each a different \texttt{for} loop, which corresponds to  a global neutral variant rate of  73\%. This neutral variant rate varies from 64\% to 92\% in \texttt{commons-collections}. This is significantly higher than any of the random \transfs analyzed previously.

\begin{table}[ht]
	\centering
	\caption{Neutral variant rate of \loopflip}
	\small{
		\begin{tabularx}{\columnwidth}{lXXXX}
			\hline
			& \#Locs & \#Trials & \#NV & NVR \\
			\hline
				commons-codec & 42 & 42 & 31 & 73\% \\
				commons-collections & 61 & 61 & 56 & 92\% \\
				commons-io & 35 & 35 & 24 & 69\% \\
				commons-lang & 227 & 227 & 146 & 64\% \\
				gson & 17 & 17 & 11 & 65\% \\
				jgit & 274 & 274 & 211 & 77\% \\
			\hline
			Total & 586 & 586 & 427 & 73\% \\
			\hline
		\end{tabularx}
	}
	\label{tab:loop-flip-results}
\end{table}

The high neutral variant rate of \loopflip  can be explained by the fact that in many cases this transformation processes loops in which there are no loop-carried dependencies \cite{devan2014review} (e.g., \autoref{lst:ex_loopflip_transf}). 
Meanwhile we can also note that both the number of candidates and the neutral variant rate vary widely from one project to another. 
This can be explained by different usages of loops in different projects. For example, if a project uses \texttt{forEach} loops more often than \texttt{for} loops, then  the number of candidates for our transformation decreases. Also,  \texttt{for} loops  are used for different purposes:  in some cases this control structure is used to apply the same computation to   $n$ elements that are independent of each other, whereas in other cases it is used to sequence of computations in which each action depends  on the previous one. In the former case, the order of the loop iteration does not matter, while in the latter case, flipping loop order is very likely to modify the global behavior.

\begin{figure}[ht]
	\begin{lstlisting}[caption={Altering an \texttt{ordered loop} in ReflectiveTypeAdapterFactory:140 (gson)},label={lst:orderedLoop},numbers=left,escapechar=@,firstnumber=140]
    List<String> fieldNames = getFieldNames(field);
    BoundField previous = null;
    Map<String, BoundField> result = new LinkedHashMap<String, BoundField>();
    [...]
@\textcolor{red}{-}@   @\textcolor{red}{for (int i = 0; i < fieldNames.size(); ++i) \{}@
@\textcolor{green}{+}@   @\textcolor{green}{for (int i = (fieldNames.size()) - 1; i >= 0; --i) \{}@
      String name = fieldNames.get(i);
      if (i != 0) serialize = false; // only serialize the default name
      BoundField boundField = createBoundField(context, field, name,
      TypeToken.get(fieldType), serialize, deserialize);
      BoundField replaced = result.put(name, boundField);
      if (previous == null) previous = replaced;
    }
	\end{lstlisting}
\end{figure}

For example, \autoref{lst:orderedLoop} illustrates a \loopflip transformation that yields a variant that is not neutral.  This case is similar to the one discussed on \autoref{lst:orderedMap}: when changing the iteration order, the \texttt{result} map is filled in a different order than in the original case. Consequently, the behavior of the method changes, which does not correspond to the expectation of the callers and eventually fails some test cases.

\newpage
\subsubsection{\modified{Research Question 5 Conclusion}}

In this section we have leveraged the observations made with generic, random \transfs, in order to design three new transformations that target code regions which are very likely plastic. 
When designing these transformations, we also increased the amount of static analysis performed by the transformation, leveraging the strong type system of Java. 
Overall, these design decisions aim at focusing the search on spaces of program variants with high densities of neutral variants.
The results confirm these higher densities, with neutral variant rates of 66\% (\addmi), 58\% (\swap), 73\% (\loopflip) that are significantly higher than the rates with generic, random transformations 23\% overall \nmodified{(p-value $< 0.001$ for each of the three \texttt{Wilcoxon rank sum test})}.

Beyond the results and observations made with these three transformations, the experiments reported here are very encouraging to explore the `grey' zone that exists between sound and semantic preserving transformations at one extreme and random, generic highly \transfs on the other extreme. We believe that in-depth knowledge about the nature of plastic code regions, combined with static code analysis is essential to design transformations that explore spaces of program variants that are behaviorally diverse, while limiting the amount of resources required to explore these spaces. 

\begin{framed}
	Answer to RQ5. \ctransfs targeted at specific plastic code regions 
are  significantly more effective than random transformations at synthesizing program variants, which exhibit visible behavior diversity and are equivalent modulo test suite. This RQ has explored three targeted \transfs that yield 66\%, 58\%, 73\% neutral variants.
\end{framed}

\section{Discussion}
\label{sec:discussion}

Our journey among the different factors that influence the synthesis of \emph{neutral program variants} has shed the light on several key findings. 
We have observed that many neutral variants result from the very specific combinations of one  \transf on one specific type of language structure. For example, the \delete transformation in ``invocation'' nodes is surprisingly effective at synthesizing neutral variants, while it performs very poorly on ``loop''. Similarly, the \add transformation is very effective with ``try'' nodes, but is very bad with ``return''. 

These observations are novel and very interesting to design \transfs in future work. Yet, we believe that the most intriguing findings of our work relate to regions of the code that are plastic by nature, and not, by chance, because of one specific transformation.

The functional contract of a code region is what ultimately determines if a variant of that region is neutral or not. 
Such a contract defines a set of properties about the inputs and outputs of the code region, as well as state invariants for that region. 
Consequently, a contract can be more or less restrictive on the behaviors that implement the contract. 
Our empirical inquiry of \transfs has revealed that some contracts define loose expectations about the behavior of a code region. In turn, these code regions are more plastic than other parts.

Here are three examples of code regions with loose contracts:
\begin{itemize}
	\item the contract of a \texttt{hash function} ( e.g. the one of \autoref{lst:coll-hash}) loosely specifies the returned value:\footnote{Oracle's documentation (java 8): \url{https://docs.oracle.com/javase/8/docs/api/java/lang/Object.html\#hashCode--}} it only enforces the result to be a deterministic integer only depending on information used in \texttt{equals}. In addition, a weak requirement is that this method should avoid collision. This means any transformation, which side effect is to change the return in a deterministic way, yields a variant that fulfills the contract, even if changing the likelihood of collision impacts performance. 
    
    \item the contract over some data ordering. For example, data structures that do not impose an order on their elements, or loops with no loop-carried dependence are code regions that have a loose contract. These regions tolerate many types of transformations that change order, for example, \loopflip or \swap in case where an ordered collection is replaced by another a non-ordered one.

	\item optional functionalities (e.g., optimization code). The elective nature of these code regions make them naturally loosely specified. These functionalities are called by other functions, and the functional contract is defined on these other functions, not on the optional ones. All transformations that remove or modify the optional functionality produce program variant that is very likely to satisfy the contract.
\end{itemize}

In this work, we have used unit test suites as proxies for functional contracts. As discussed in \autoref{sec:taxonomy}, this might lead to false positives (variants considered neutral modulo the test suite, but that happen to be buggy variants). Yet, in many cases, this also allowed us to spot  \emph{inherently plastic} code regions that are prone to several \transfs, which can synthesize  more neutral variants.

\newpage
\section{Threats to validity}
\label{sec:threats}

We performed a large scale experiment in a relatively unexplored domain: the characterization of plastic code regions.
We now present the threats to validity. 

While we aim at analyzing code plasticity, we actually  measure the  rate of neutral variants produced by specific \transfs. This can raise a threat to the construct validity of our study, with respect to two concerns: i) the limitation of plasticity to a given transformation, ii) the confinement of changes only to the source code but not to the behavior.
We mitigate the first concern through the manual analysis in our answer to RQ4 that emphasizes the presence of real code plasticity and not only plasticity related to a given transformation. To mitigate the second, we analyzed, in RQ5's answer, the execution traces proving actual, observable differences in execution.

Our findings might not generalize to all types of applications. Depending on the type of applications and the quality of their test suite, the obtained results could change. To minimize the impact of this threat, we selected open source frameworks and libraries because of their popularity, their longevity and the very high quality of their test suites.
In addition, we provided an explicit analysis of the impact of tests on the neutral variant rate of transformations in \autoref{sec:test}.

Finally, our large scale experiments rely on a complex tool chain, which integrates code transformation, instrumentation, trace analysis and statistical analysis. 
We also rely on the Grid5000 grid infrastructure to run millions of transformations.
We did extensive testing of our code transformation infrastructure, built on top of the Spoon framework that has been developed, tested and maintained for more than 10 years.
However, as for any large scale experimental infrastructure, there are surely bugs in this software. We hope that they only change marginal quantitative results, and not the qualitative essence of our findings. Our infrastructure is publicly available on Github.\footnote{\scriptsize \url{https://github.com/castor-software/journey-paper-replication}}

\newpage
\section{Related work}
\label{sec:related}

Our work is related to the seminal work, which analyze the capacity of software at yielding useful variants under \transfs. It is also related to work that exploit \transf (either random or targeted) to improve software. 
Here, we discuss the key work in these areas, as well as the novelty of our work.

\vspace{-0.5cm}
\subsection{Plasticity of software}
\vspace{-0.2cm}

The work on \emph{mutational robustness} by Schulte and colleagues \cite{Schulte13} is a key inspiration for our own work. 
These authors explore the ability of software to be transformed under random copy, deletion and swap of AST nodes. Their experiments on 22 small to medium C programs (30 to 60 K lines of code) show that 30 \% of the transformations yield variants that are equivalent to the original, modulo the test suite. They call this property of software \emph{mutational robustness}. More recently, this research group demonstrate that the interaction of several neutral mutations can lead a program to exhibit new positive behavior such as passing an additional test. They call this phenomenon \emph{positive epistasis}  \cite{Renzullo2018}. \modified{Other work has since confirmed the existence of mutational robustness\cite{haraldsson2017,langdon2017}}

Our RQ1 can be considered as a conceptual replication \cite{shull_role_2008} of the work by Schulte and colleague. 
Our results mitigate two threats to the validity of Schulte's results: our methodology mitigates internal threats, by using another tool to perform \transfs, and our experiment mitigates external threats by transforming Java programs (instead of C). Similarly to Schulte, we conclude ``that mutational robustness is an inherent property of software''. 
Yet, our study also provides completely novel insights about the language constructs and the code areas that support mutational robustness (we call them \emph{plastic code regions}) and about the effectiveness of targeted transformations to maximize the synthesis of neutral variants.

Recently, Danglot and colleagues have also explored the capacity of software and absorbing state transformations \cite{DanglotPBM16}. They explore \emph{correctness attraction}: the extent to which programs can still produce correct results under runtime state perturbations. In that study the authors rely on a perfect oracle to asses the correctness of outputs, and they  observe that many perturbations do not break the correctness in ten subject programs.
Our work also shows that program variants can have different traces and still deliver equivalent results (modulo the test suite). Yet, we rely on different transformations and we analyze in-depth the nature of the code regions that can yield neutral variants.

Our work extends the body of knowledge about forgiving code regions \cite{rinard2012}.
In particular, we find regions characterized by ``plastic specifications'', i.e. regions  which are governed by a very open yet strong contract. For instance, the only correctness contract of a hashing function is to be deterministic. On the one hand this is a strong contract. On the other hand, this is very open: many variants of a hashing function are valid, and consequently, many modifications in the code result in valid hashing functions.

Some recent work investigate a specific form of software plasticity, referred to as \emph{redundancy} \cite{Carzaniga2015,gabel2010study,suzuki2017exploratory}. This work consider that a code fragment is redundant with another fragment, in a specific context, if in that context, both fragments lead a program from a given state to an equivalent one through a different series of intermediate state. 
This is very close to neutral variants, which have diverse visible behavior and yet satisfy the same properties as assessed by the test suite. The key difference between our work is that we investigate \transfs to synthesize neutral variants, i.e. increase redundancy, whereas they analyze redundancy that naturally occurs in software systems.

\vspace{-0.5cm}
\subsection{Exploiting software plasticity}
\vspace{-0.2cm}

\modified{Genetic improvement \cite{petke2017genetic} is an area of search-based software engineering \cite{harman2001search}, which consists in automatically and incrementally generating variants of an existing program in order to either improve non-functional properties such as resource consumption or execution time, or functional ones (e.g. automatic repair).} All variants should pass the test suite of the original program.
Existing work in this domain rely on random \transfs to search for program variants: Schulte and colleagues \cite{schulte2014} exploit mutational robustness to reduce energy consumption; \modified{Langdon \textit{et al} \cite{langdon2017software} add, delete, replace lines in C, C++, CUDA program sources to improve performance;  Cody-Kenny \textit{et al}. \cite{perfLoc} add, delete, replace AST nodes, to profile program performance;} L{\'{o}}pez and colleagues \cite{lopez18} explore program mutations to optimize source code. \modified{Manotas and al\cite{Manotas} replace java collections to optimize energy consumption.}
All this work leverage the existence of code plasticity, and the performance of the search process can be improved with targeted \transfs. In particular, our results with the \swap transformation, show that changing library is very effective to generate neutral variants, and this transformation is a key enabler to improve performance \cite{DBLP:journals/corr/BasiosLWKLB17}.

Software diversification \cite{baudry2015} is the field concerned with the automatic synthesis of program variants for dependability. Existing work in this area also intensively exploit software plasticity and \transfs: Feldt \cite{Feldt98} was among the first to  use genetic programming to generate multiple versions of a program to have failure diversity; we relied on random transformations to synthesize diverse implementations of Java programs \cite{allier14,Baudry14}; recent work on composite diversification \cite{wang2017composite}, investigate the opportunity to combine multiple security oriented transformation techniques. This work can benefit from our findings about targeted \transfs, which introduce important behavior changes (in particular the \swap transformation), while maximizing the chances of preserving the core functionality.

Shacham and colleagues \cite{Shacham} and, more recently, Basios and colleagues \cite{DBLP:journals/corr/BasiosLWKLB17} investigate source code transformations to replace libraries and data structures, in a similar was as the \swap transformation.  This corroborates the idea of a certain plasticity around these data structures, and the notion of interface.

\section{Conclusion}
\label{sec:conclusion}

The existence of neutral program variants and the ability to generate large quantities of such variants are essential foundations for automatic software improvement. Our work contributes to these foundations with novel empirical facts about neutral variants and with actionable transformations to synthesize such variants. Our empirical analysis explores the space of neutral variants of Java programs, focusing on 6 large open source projects, from different domains. We generated 98225 variants that compile for these projects, through \transfs, and 23445 were neutral variants, i.e., more than 20\% of the variants run correctly and pass the same test suite as the original. A detailed analysis of these neutral variants revealed that some language constructs are more likely to be plastic than others to the synthesis of neutral variants (for example, method invocations) and also that some code regions have specific roles that make them plastic (for example, optimization code).

The actionable contribution of our work comes in the form of three novel \transfs for Java programs. We have designed these transformations to target specific code regions that appear more prone to neutral variant synthesis. Our experiments show that these transformations perform significantly better than generic ones: 60\% (\addmi), 58\% (\swap), 73\% (\loopflip) instead of 23,9\%.

One key insight from the series of experiments reported in this work is that some code regions are \emph{inherently} plastic. These code regions are naturally prone to behavioral variations that preserve the global functionality. These regions include code that has a plastic specification (e.g., hash function); optional functionality (e.g., optimization code) or regions that can be naturally reordered (e.g., loops with no loop-carried dependence). In our future work, we wish to leverage this insight about the deep nature of large programs to develop techniques that can generate vast amounts of software diversity for obfuscation \cite{harrandsoftware} and moving target defenses \cite{Okhravi14moving}.


\appendix
\section{Appendix}
\label{sec:appendix}

\subsection{Add method invocation}

The following section details the transformation Add method invocation. Listing \ref{lst:addmi-target} describes the subset of the java language targeted, and what follows describes the transformation's behavior.

\lstset{language=Perl,escapeinside={(*@}{@*)},label={lst:addmi-target}}
\begin{lstlisting}
<class> ::= (<modifier>)*
'class' <identifier> (ref)* 
'{' (<class_member>)* '}'

<class_member> ::= <attribute> | 
<other> | <method>

<method> ::= (<modifier>)* (return_type) 
<identifier> '(' 
(<type> <identifier>)* ')' '{' 
(<statement>)* '}'

<statement> ::= <variable_declaration> |
<block> | <other>

<block> ::= (<block_header>) '{' 
(<statement>)* '}'
\end{lstlisting}

$P = \{\text{Packages}\},$

$C(p) = \{\text{Classes of } p \| p \in P\},$

$M(c) = \{\text{methods in class } c\},$ 

$S(m) = \{\text{statements in method } m \text{'s body}\},$

$LV(s) = \{\text{Local variables up to } s\},$ 

$Pa(m) = \{\text{Parameter of method } m\},$ 

$A(c) = \{\text{Attributes of class } c\},$ 

$As(c) = \{a \in A(c) \| static(a)\},$ 

Let $p \in P, c \in C(p), m \in M(c), s \in S(m)$

$V(c,m,s) = LV(s) \cup Pa(m) \cup A(c) \cup \{ this \},$
a set of accessible variable from $s$

$Vs(c,m,s) = LV(s) \cup Pa(m) \cup As(c),$
a set of accessible variable from $s$

$\mathcal{M}(p,c,m,s) = \{ m' \in c\} \cup \{ m' \in c' \| \forall c' \in p \land \lnot private(m')\} \cup \{m' \in c' \| \forall p' \in P, \forall c' \in p' \land public(m')\}$

$\mathcal{M}_{a}(p,c,m,s) = \{ m' \| static(m') \lor ClassOf(m') \in TypeOf(LV(c,m,s))\}$

Let $m' \in \mathcal{M}_{a}$

\inference[$\text{Class}$]{
	c\text{ follows } ... m ...
}{
	c -> ... Well;m ...
}

\vspace{1em}

\inference[$\text{Well}_{static \land \lnot void}$]{
	\text{static}(m) \; \; \land \; \; \text{TypeOf}(m') \neq void
}{
	Well -> \text{'public static'} \; \text{TypeOf}(m') \; \text{wellID} 
}

\vspace{1em}

\inference[$\text{Well}_{\lnot static \land \lnot void}$]{
	\lnot \;\text{static}(m) \; \land \; \text{TypeOf}(m') \neq void
}{
	Well -> \text{'public'} \;  \text{TypeOf}(m') \; \text{wellID} 
}

\vspace{1em}

\inference[$\text{Well}_{void}$]{
	\text{TypeOf}(m') = void
}{
	Well -> SKIP
}

\vspace{1em}

\inference[$\text{Method}$]{
	m\text{ follows } ... s ...
}{
	m -> ...  \text{'try \{'} Well \; Ta \; Call; \text{'\} catch (Exception ' eId ') \{\}'} s ...
}
										
\inference[$\text{We}$]{
	\text{TypeOf}(m') \neq void
}
{
	We -> wellID = 
}

\vspace{1em}

\inference[$\text{We}_{void}$]{
	\text{TypeOf}(m') = void
}{
	We -> SKIP
}

\vspace{1em}

\inference[$\text{Target}_{static}$]{
	\text{static}(m')
}{
	Ta -> SKIP
}

\vspace{1em}

\inference[$\text{Target}$]{
	\lnot \; \text{static}(m')
}{
	Ta -> targetID.
}

\vspace{1em}

\inference[$\text{Call}$]{
}{
	Call -> QN(m')(params)
}

\subsection{SwapSubtype}
\label{sec:swap-appendix}
The following section details the behavior of the SwapSubtype transformation, the subset of java targeted \ref{lst:swap-target}, and how it modified it.

\lstset{language=Perl,escapeinside={(*@}{@*)},label={lst:swap-target}}
\begin{lstlisting}
<affectation> ::= <ls> '=' <rs>
<ls> ::=  (<interface> ('<' <type> '>')?)? <identifier>
<rs> ::= 'new' <concrete_class_constructor> ('<' <type> '>')? '(' <param_list> ')'
<param_list> ::= <> | <param> | <param> ',' <param_list>
\label{lst:col-assign}
\end{lstlisting}

$ I = \{\text{Interfaces}\}$

$ T = \{\text{Types}\}$

$ C(t) = \{\text{Constructor of }t\}$

$ T(i) = \{t \in T \| t \text{ implements } i\}$

Let $i \in I, t_{1},t_{2} \in T(i)^2$ such as $t_{1} \neq t_{2}$

\inference[$\text{Affectation}$]{
	\text{Aff}(I,id,t_{1},params) \land \exists c \in C(t_{2}) 
}
{
	\text{Aff}(I,id,t_{1},params) -> I id = 'new' t_{2} '(' params ')'
}

The following sections list the different interfaces targeted by our implementation of the transformation, and for each interface the different classes implementing these interfaces used interchangeably.

\subsubsection{java.util.SortedSet}
\label{sec:col-list}
\begin{itemize}
\item java.util.concurrent.ConcurrentSkipListSet
\item java.util.TreeSet
\end{itemize}
\subsubsection{java.util.concurrent.BlockingDeque}
\begin{itemize}
\item java.util.concurrent.LinkedBlockingDeque
\end{itemize}
\subsubsection{java.util.Collection}
\begin{itemize}
\item java.util.concurrent.LinkedTransferQueue
\item java.util.concurrent.SynchronousQueue
\item java.util.PriorityQueue
\item java.util.concurrent.CopyOnWriteArraySet
\item java.util.concurrent.LinkedBlockingQueue
\item java.util.TreeSet
\item java.util.concurrent.ConcurrentLinkedDeque
\item java.util.Stack
\item java.util.concurrent.PriorityBlockingQueue
\item java.util.ArrayList
\item java.util.HashSet
\item java.util.concurrent.ArrayBlockingQueue
\item java.util.Vector
\item java.util.concurrent.ConcurrentSkipListSet
\item java.util.concurrent.LinkedBlockingDeque
\item java.util.concurrent.DelayQueue
\item java.util.ArrayDeque
\item java.util.LinkedList
\item java.util.LinkedHashSet
\item java.util.concurrent.ConcurrentLinkedQueue
\item java.util.concurrent.CopyOnWriteArrayList
\end{itemize}
\subsubsection{java.util.concurrent.ConcurrentNavigableMap}
\begin{itemize}
\item java.util.concurrent.ConcurrentSkipListMap
\end{itemize}
\subsubsection{java.util.Set}
\begin{itemize}
\item java.util.HashSet
\item gnu.trove.set.hash.THashSet
\item java.util.concurrent.ConcurrentSkipListSet
\item org.apache.commons.collections4.set.ListOrderedSet
\item java.util.concurrent.CopyOnWriteArraySet
\item java.util.TreeSet
\item java.util.LinkedHashSet
\item gnu.trove.set.hash.TCustomHashSet
\end{itemize}
\subsubsection{java.util.concurrent.BlockingQueue}
\begin{itemize}
\item java.util.concurrent.ArrayBlockingQueue
\item java.util.concurrent.LinkedTransferQueue
\item java.util.concurrent.SynchronousQueue
\item java.util.concurrent.LinkedBlockingDeque
\item java.util.concurrent.DelayQueue
\item java.util.concurrent.LinkedBlockingQueue
\item java.util.concurrent.PriorityBlockingQueue
\end{itemize}
\subsubsection{java.util.NavigableSet}
\begin{itemize}
\item java.util.concurrent.ConcurrentSkipListSet
\item java.util.TreeSet
\end{itemize}
\subsubsection{java.util.Deque}
\begin{itemize}
\item java.util.concurrent.LinkedBlockingDeque
\item java.util.ArrayDeque
\item java.util.LinkedList
\item java.util.concurrent.ConcurrentLinkedDeque
\end{itemize}
\subsubsection{java.util.concurrent.TransferQueue}
\begin{itemize}
\item java.util.concurrent.LinkedTransferQueue
\end{itemize}
\subsubsection{java.util.NavigableMap}
\begin{itemize}
\item java.util.concurrent.ConcurrentSkipListMap
\item java.util.TreeMap
\end{itemize}
\subsubsection{java.util.concurrent.ConcurrentMap}
\begin{itemize}
\item java.util.concurrent.ConcurrentSkipListMap
\item java.util.concurrent.ConcurrentHashMap
\end{itemize}
\subsubsection{java.util.List}
\begin{itemize}
\item org.apache.commons.collections4.list.TreeList
\item java.util.Vector
\item org.apache.commons.collections4.list.NodeCachingLinkedList
\item org.apache.commons.collections4.list.CursorableLinkedList
\item java.util.LinkedList
\item org.apache.commons.collections4.list.GrowthList
\item java.util.Stack
\item java.util.ArrayList
\item java.util.concurrent.CopyOnWriteArrayList
\item org.apache.commons.collections4.ArrayStack
\end{itemize}
\subsubsection{java.util.Map}
\begin{itemize}
\item org.apache.commons.collections4.map.SingletonMap
\item org.apache.commons.collections4.map.Flat3Map
\item org.apache.commons.collections4.map.LinkedMap
\item java.util.concurrent.ConcurrentHashMap
\item org.apache.commons.collections4.map.LRUMap
\item org.apache.commons.collections4.map.ListOrderedMap
\item java.util.HashMap
\item org.apache.commons.collections4.map.HashedMap
\item org.apache.commons.collections4.map.ReferenceMap
\item org.apache.commons.collections4.map.CaseInsensitiveMap
\item gnu.trove.map.hash.TCustomHashMap
\item java.util.LinkedHashMap
\item org.apache.commons.collections4.map.PassiveExpiringMap
\item java.util.concurrent.ConcurrentSkipListMap
\item org.apache.commons.collections4.map.StaticBucketMap
\item java.util.TreeMap
\item gnu.trove.map.hash.THashMap
\item java.util.Hashtable
\item java.util.WeakHashMap
\item org.apache.commons.collections4.map.ReferenceIdentityMap
\end{itemize}
\subsubsection{java.util.Iterable}
\begin{itemize}
\item java.util.concurrent.LinkedTransferQueue
\item java.util.concurrent.SynchronousQueue
\item java.util.PriorityQueue
\item java.util.concurrent.CopyOnWriteArraySet
\item java.util.concurrent.LinkedBlockingQueue
\item java.util.TreeSet
\item java.util.concurrent.ConcurrentLinkedDeque
\item java.util.Stack
\item java.util.concurrent.PriorityBlockingQueue
\item java.util.ArrayList
\item java.util.HashSet
\item java.util.concurrent.ArrayBlockingQueue
\item java.util.Vector
\item java.util.concurrent.ConcurrentSkipListSet
\item java.util.concurrent.LinkedBlockingDeque
\item java.util.concurrent.DelayQueue
\item java.util.ArrayDeque
\item java.util.LinkedList
\item java.util.LinkedHashSet
\item java.util.concurrent.ConcurrentLinkedQueue
\item java.util.concurrent.CopyOnWriteArrayList
\end{itemize}
\subsubsection{java.util.Queue}
\begin{itemize}
\item java.util.concurrent.LinkedTransferQueue
\item java.util.concurrent.SynchronousQueue
\item java.util.PriorityQueue
\item java.util.concurrent.LinkedBlockingQueue
\item java.util.concurrent.ConcurrentLinkedDeque
\item java.util.concurrent.PriorityBlockingQueue
\item java.util.concurrent.ArrayBlockingQueue
\item org.apache.commons.collections4.queue.CircularFifoQueue
\item java.util.concurrent.LinkedBlockingDeque
\item java.util.concurrent.DelayQueue
\item java.util.ArrayDeque
\item java.util.LinkedList
\item java.util.concurrent.ConcurrentLinkedQueue
\end{itemize}
\subsubsection{java.util.SortedMap}
\begin{itemize}
\item java.util.concurrent.ConcurrentSkipListMap
\item java.util.TreeMap
\end{itemize}

\subsection{Loopflip}
\label{apx:loopflip}

\lstset{escapeinside={(*@}{@*)},label={lst:loopflip-target}}
\begin{lstlisting}
<loop> ::= 'for(' <initialization> ';'
<condition> ';' <update> ')' '{' 
(<statement>)* '}'

<initialization> ::= <identifier> 
'=' <expression>

<condition> ::= <identifier> 
<binary_operator> <expression>

<binary_operator> ::= '<' |
'<=' | '>' | '>='

<update> ::= <identifier> '=' 
<identifier> <operator> <expression>

<operator> ::= '+' | '(*@-@*)'
\end{lstlisting}
We extends update statements such as \texttt{i++} into \texttt{i = i + 1} and \texttt{i -= 2} into \texttt{i = i - 2}

$\text{comp} \in \{<,>,\geq,\leq\},\text{op} \in \{+,-\}$

$\overline{a}, \forall a \in  \{<,>,\geq,\leq\} \cup \{+,-\}
\left \{
\begin{array}{rcl}
>,\geq  & \mapsto & \leq \\
<,\leq   & \mapsto & \geq \\
+ & \mapsto & - \\
- & \mapsto & +
\end{array}
\right .
$

\inference[$\text{For}_{L}$]{
	\text{comp} \in \{\geq,\leq\} \; \land \; |i_{end} - i_{0}| \equiv 0 \pmod p
	}
	{
		(\text{For}_{L}(i=i_{0} i \text{ comp } i_{end} i = i \text{ op } p)-> (\text{For}_{L}(i=i_{end} i \text{ }\overline{\text{comp}}\text{ } i_{0} i = i \text{ }\overline{\text{op}}\text{ } p)
		}
		
		\inference[$\text{For}_{L}$]{
			\text{comp} \not\in \{\geq,\leq\} \; \lor \; \lnot(|i_{end} - i_{0}| \equiv 0 \pmod p)
			}
			{
				(\text{For}_{L}(i=i_{0} i \text{ comp } i_{end} i = i \text{ op } p) -> (\text{For}_{L}(i=i_{end} \; \overline{\text{op}} \;(|i_{end} - i_{0}| \pmod p)  i \text{ }\overline{\text{comp}}\text{ } i_{0} i = i \text{ }\overline{\text{op}}\text{ } p)
				}

\end{document}